%% file: main.tex
\documentclass[letterpaper,twocolumn,10pt]{article}
\input{latex_config}

\begin{document}
% ----------------------------------------------------

\begin{textblock}{12}(2,1)
\centering
To Appear in the 43rd IEEE Symposium on Security and Privacy, May 22–26, 2022.
\end{textblock}

% ----------------------------------------------------
\title{\Large \bf Model Stealing Attacks Against Inductive Graph Neural Networks}
% ----------------------------------------------------

\date{}

\author{
Yun Shen\textsuperscript{1}\thanks{The first two authors made equal contributions.}\ \ \
Xinlei He\textsuperscript{2}\textsuperscript{\textcolor{blue!60!green}{$\ast$}}\ \ \
Yufei Han\textsuperscript{3}\ \ \
Yang Zhang\textsuperscript{2}
\\
\\
\textsuperscript{1}\textit{Norton Research Group}\ \ \ 
\textsuperscript{2}\textit{CISPA Helmholtz Center for Information Security}\ \ \
\textsuperscript{3}\textit{INRIA}\ \ \
}

\maketitle

% ----------------------------------------------------
\begin{abstract}
% ----------------------------------------------------

Many real-world data come in the form of graphs.
Graph neural networks (GNNs), a new family of machine learning (ML) models, have been proposed to fully leverage graph data to build powerful applications.
In particular, the inductive GNNs, which can generalize to unseen data, become mainstream in this direction.
Machine learning models have shown great potential in various tasks and have been deployed in many real-world scenarios.
To train a good model, a large amount of data as well as computational resources are needed, leading to valuable intellectual property.
Previous research has shown that ML models are prone to model stealing attacks, which aim to steal the functionality of the target models.
However, most of them focus on the models trained with images and texts.
On the other hand, little attention has been paid to models trained with graph data, i.e., GNNs.
In this paper, we fill the gap by proposing the first model stealing attacks against inductive GNNs.
We systematically define the threat model and propose six attacks based on the adversary's background knowledge and the responses of the target models.
Our evaluation on six benchmark datasets shows that the proposed model stealing attacks against GNNs achieve promising performance.\footnote{Our code is available at \url{https://github.com/xinleihe/GNNStealing}.}

% ----------------------------------------------------
\end{abstract}
% ----------------------------------------------------

% ----------------------------------------------------
\section{Introduction}
\label{sec:introduction}
% ----------------------------------------------------

Many real-world data come in the form of graphs, such as molecular graphs~\cite{KMBPR16} and social networks~\cite{S17}.
The graph-structured data contains nodes with features and edges that represent the relationship between them.
To fully unleash the potential of graph data, a new family of machine learning (ML) models, namely graph neural networks (GNNs), has been proposed~\cite{GF18,ZCZYLS18,LRKAK19,ZCZ20,WPCLZY20}.
Compared to classical machine learning models, e.g., convolutional neural networks (CNNs) and recurrent neural networks (RNNs), which are designed to process images and texts, GNNs offer state-of-the-art performance by taking both node features as well as graph structures into consideration.

Prior work unveiled that machine learning models are vulnerable to \emph{model stealing attacks} ~\cite{TZJRR16,OSF19,JCBKP20,WG18,OASF18}, where an adversary with query access to a target model can steal its parameters or functionality.
Concretely, the adversary first crafts a number of queries as the input to the target model's API and obtains the corresponding outputs.
Then, a local surrogate model is trained on the paired data (input, output).
As such, the surrogate model may not only violate the intellectual property of the target model but also serve as a stepping stone for further attacks like membership inference~\cite{SSSS17,SZHBFB19,SDSOJ19,JSBZG19,CYZF20,CTCP21,CTWJHLRBSEOR20,HZ21,LZ21,LWHSZBCFZ22} and adversarial examples~\cite{GSS15,PMWJS16,PMGJCS17,WWTDLZ19,CW17,TKPGBM17}.
Notably, most of the current efforts on model stealing attacks concentrate on ML models with images and text data~\cite{TZJRR16,OSF19,OASF18,WG18,KTPPI20}. 
On the other hand, the potential model stealing risks of GNNs have been largely understudied.

There exists some preliminary work on model stealing attacks against GNNs~\cite{DR19,WYPY20}. 
They focus on \textit{transductive} GNNs and assume that the attackers have access to the training process of the target model, in which the training and query graphs are used to train the target model.
As such, GNN model stealing attacks in a transductive setting are unrealistic.

In this paper, we concentrate on a more realistic and popularly deployed GNN setting, i.e., \textit{inductive} GNNs, which can generalize well to unseen nodes~\cite{HYL17,VCCRLB18,XHLJ19}.
In this setting, the adversary only queries the target model via remotely  
accessible  API.  
They do not tamper with the training process of the target model. 
Note that in this paper, we focus on node classification tasks.

When an adversary uses a \emph{query graph} (i.e., a graph induced by a set of query nodes) to query the target model, they could face different types of \emph{responses} from the target model, ranging from the posterior distribution over the possible labels of query nodes (called predicted posterior probability) to 2-dimensional t-SNE~\cite{MH08} projection (for graph visualization).
Therefore, it is important to summarize a complete taxonomy of the threats for model stealing attacks against GNNs.
Also, it is challenging to design a general attack methodology that can be applied to different attack scenarios.
Moreover, in reality, the graph structural information of the query graph can be missing. 
It is inevitably harder for the adversary to launch attacks against GNNs.

To tackle these challenges, we make the following contributions in this paper.
We first systematically define the threat model of model stealing attacks against inductive GNNs by categorizing the adversary's knowledge into two dimensions, i.e., query graph and target model's response (see \autoref{sec:threat_model}).
Concretely, we assume that the adversary has a query graph that contains a number of nodes and their features.
The node features come from the same distribution of the graph used to train the target model. 
However, the graph structure of the query graph may be missing (i.e., the edges connecting the nodes may not be available). 
Regarding the target model's response, we consider three cases, i.e., \emph{the predicted posterior probability}, \emph{the node embedding vector}, or the 2-dimensional \emph{t-SNE projection}.
In turn, we have six attack scenarios under our threat model.

We propose two types of attacks, i.e., Type I and Type II, based on the information provided by the query graph (see \autoref{sec:attack_taxonomy}).
Each type has three variants depending on the target model's responses.
We design a general attack framework that can be applied to all the scenarios.
Concretely, the framework is assembled with two major components.
The first component is used to learn the discrete graph structure if the structural information is not available in the query graph.
Then, the second component builds a surrogate model by jointly learning from the nodes' features and the response of the target model.

Ideally, the surrogate model should achieve both high accuracy and high fidelity whereby accuracy measures the prediction correctness~\cite{TZJRR16,PMGJCS17} and fidelity measures the prediction agreement between the target model and the surrogate model~\cite{JCBKP20,KTPPI20}.
We evaluate all our attacks on three popular inductive GNN models including GraphSAGE~\cite{HYL17}, Graph Attention Network (GAT)~\cite{VCCRLB18}, and Graph Isomorphism Network (GIN)~\cite{XHLJ19} with six benchmark datasets, i.e., DBLP~\cite{PWZZW16}, Pubmed~\cite{SNBGGE08}, Citeseer Full~\cite{GBL98}, Coauthor Physics~\cite{SMBG18}, ACM~\cite{WJSWYCY19}, and Amazon Co-purchase Network for Photos~\cite{MTSH15}.
Extensive experiments demonstrate that our model stealing attacks consistently achieve strong performance in both types of attacks. 
For instance, when the target model is GIN trained on the Pubmed dataset and the response is embeddings, our Type I attack achieves a 0.877 accuracy score and a 0.906 fidelity score (see \autoref{sec:type1_performance}).
In particular, when the aforementioned target model's response is the t-SNE projection, our attack still achieves strong performance with a 0.823 accuracy score and a 0.846 fidelity score. 
Moreover, we empirically demonstrate that even without graph structure information, i.e., Type II attacks, the adversary could still launch effective attacks, extracting high-accuracy and high-fidelity surrogate models (see \autoref{sec:type2_performance}).
This further demonstrates the severe model stealing risks of GNN models.

In summary, we make the following contributions.

\begin{itemize}
\item Our work is the first research effort to perform model stealing attacks against inductive GNNs.
\item We systematically define the threat model to characterize an adversary's background knowledge along two dimensions.
Moreover, we propose six different attack scenarios based on the adversary's different background knowledge. 
\item Extensive evaluation on three popular inductive GNN models and six benchmark graph datasets demonstrates the efficacy of our attacks.
\end{itemize}

% ----------------------------------------------------
\section{Background}
\label{sec:background}
% ----------------------------------------------------

% ----------------------------------------------------
\subsection{Notations}
\label{sec:notations}
% ----------------------------------------------------

We define a labelled, undirected, unweighted, attributed graph as $\mathbf{G}=(\mathbf{V}, \mathbf{E}, \mathbf{X}, \mathbf{C})$, where $\mathbf{V}=\{v_1, v_2, ..., v_n\}$ denotes the set of nodes, $\mathbf{E} \subseteq \{(v, u)| v, u \in \mathbf{V}\}$ denotes the set of edges, $\mathbf{x}_i \in \mathbf{X}$ denotes the feature of node $v_i$, and one-hot vectors $\mathbf{c}_i \in \mathbf{C}$ denotes the label of node $v_i$. 
We denote $\mathbf{A} \in \{0,1\} ^ {n \times n}$ as the adjacency matrix, where $\mathbf{A}_{vu} = 1, \forall (v,u) \in \mathbf{E}$. 
As such, $\mathbf{G}$ can also be represented as $\mathbf{G}=(\mathbf{A}, \mathbf{X}, \mathbf{C})$.
The original graph used to train a GNN model is denoted as $\mathbf{G}_O$ (training graph).
We use $\mathcal{N}^l(v)$ to denote $l$-hop neighborhood of $v$, and $\mathbf{G}_v^l$ to denote the subgraph induced by $l$-hop neighborhood of $v$.
Besides, we denote the target GNN model as $\mathcal{M}_T$ and the surrogate GNN model as $\mathcal{M}_S$.
The notations introduced here and in the following sections are summarized in \autoref{tab:notations}.

\begin{table}[!t]
\centering
\caption{Summary of the notations used in this paper. 
We use lowercase letters to denote scalars, bold lowercase letters to denote vectors and bold uppercase letters to denote matrices. }
\scalebox{0.85}{
\begin{tabular}{l|l}
\toprule 
\textbf{Notation} & \textbf{Description} \\ \midrule
 $\mathbf{G}=(\mathbf{V}, \mathbf{E}, \mathbf{X}, \mathbf{C})$   & graph   \\ 
 $v,u \in V$ &  node \\
 $n=|\mathbf{V}|$ & number of nodes \\
 $\mathbf{A} \in \{0,1\} ^ {n \times n}$ & adjacency matrix \\ 
 $d$ & dimension of a node feature vector \\
 $b$ & dimension of a hidden node feature vector \\
 $\mathbf{c}_i \in \mathbf{C}$ & node label/class \\
 $\mathcal{N}^l(v)$ & $l$-hop neighborhood of $v$ \\
 $\mathbf{G}_v^l$ & subgraph induced by $l$-hop neighborhood of $v$ \\
 $\mathbf{X} \in \mathbb{R}^{n \times d}$ & node feature matrix \\
 $\mathbf{x}_v \in \mathbb{R}^{d}$  & feature vector of node $v$ \\ 
 $\mathbf{H} \in \mathbb{R}^{n \times b} $ & hidden state matrix \\
 $\mathbf{h}_v \in \mathbb{R}^{b} $ & hidden state of $v$ \\
 $\mathbf{G}_O$/$\mathbf{G}_Q$ & training/query graph \\ 
 $\mathbf{R}$ & query response  \\
 $\boldsymbol{\Upsilon} \in \mathbb{R}^{n \times 2}$ & 2-dimensional t-SNE projection\\ 
 $\boldsymbol{\Theta} \in \mathbb{R}^{n \times {|\mathbf{C}|}}$ & predicted posterior probability \\ 
 $\mathcal{M}_T$/$\mathcal{M}_S$ & target/surrogate GNN model \\
\bottomrule	 
\end{tabular}
}
\label{tab:notations}
\end{table}

% ----------------------------------------------------
\subsection{Preliminaries}
\label{sec:preliminaries}
% ----------------------------------------------------

\mypara{Graph Neural Networks (GNNs)}
Representation learning of graph-structured data is challenging because both graph structure and node features carry important information.
Graph Neural Networks (GNNs) provide an effective way to fuse information from network structure and node features.
Most of the GNNs follow a neighborhood aggregation strategy, where the model iteratively updates the representation of a node through message passing and aggregating representations of its neighbors. 
After $l$ iterations of aggregation, a node’s representation, denoted as $\mathbf{h}_v$, captures the structural information within its $l$-hop network neighborhood. 
In practice, a GNN contains several graph convolutional layers.
Each graph convolutional layer of a GNN model can be defined as follows:

\begin{equation}
\begin{array}{l}
\mathbf{h}_v^{l} = \mathrm{AGGREGATE} ( \mathbf{h}_v^{l-1},  \mathrm{MSG} ( \mathbf{h}_v^{l-1}, \mathbf{h}_u^{l-1})), u \in \mathcal{N}(v) 
\end{array}
\end{equation}

Note that the number of graph convolutional layers is equivalent to the $l$-hop network neighborhood that a GNN model can reach in the graph.
Once trained, the GNN can map each node to an embedding vector.
These node embeddings can be directly used for downstream machine learning tasks that can be categorized into three levels, i.e., node-level (e.g., node classification~\cite{KW17,HYL17,VCCRLB18,XHLJ19}), link-level (e.g., link prediction~\cite{PAS14,GL16,BKW17}), and graph-level (e.g., graph classification~\cite{LLK19,YYMRHL18}).

\mypara{Inductive GNN Models}
There are two settings for training the GNNs, i.e., transductive setting and inductive setting.
In the transductive setting, a GNN learns from both labelled and unlabelled nodes in a single fixed graph at the training time and predicts the labels of those unlabelled nodes once the training is done, e.g., vanilla graph convolutional network (GCN)~\cite{KW17}, DeepWalk~\cite{PAS14}, etc.
However, transductive GNN models must be retrained if new nodes are introduced to the graph.
A more popular one is the setting of inductive learning, where the learned GNN model can be generalized to the graphs that are previously unseen during the training procedure.
The reusable GNN model avoids time-consuming retraining if a graph includes more nodes or even subgraphs. 
It facilitates the real-world practices of graph data analytics. 
We therefore focus on the inductive setting in our study. 
We briefly introduce three widely used inductive GNN models below.

\begin{itemize}
\item \textit{GraphSAGE.} 
GraphSAGE proposed by Hamilton~et al.~\cite{HYL17} is the first inductive GNN model.
Inspired by the Weisfeiler-Lehman test for graph isomorphism, GraphSAGE generalizes the original GCN~\cite{KW17} into the inductive setting with different aggregation functions.
Take widely used mean aggregation operator as an example, GraphSAGE can be defined as follows:
\begin{equation}
\mathbf{h}_{v}^{l} = \mathrm{MEAN} (\mathbf{h}_{v}^{l-1} \cup \{ {\mathbf{h}_{u}^{l-1}, \forall u \in \mathcal{N}(v)} \} ) \label{eq:graphsage}
\end{equation}
\item \textit{Graph Attention Network (GAT).}
It is straightforward to observe that GraphSAGE assigns the same weight to all neighbors (i.e., $1/{\mathcal{N}(v)}$) when aggregating $v$'s neighborhood information.
However, in practice, different nodes may play different roles in the target node embedding.
Inspired by the attention mechanism in deep learning~\cite{BCB15}, Velickovic~et al.~\cite{VCCRLB18} propose GAT that leverages multi-head attention to learn different attention weights and pays more attention to the important neighborhoods.
Its aggregation function can be formulated as:
\begin{equation}
\mathbf{h}_{v}^{l} =    \displaystyle\Big\Vert_{z=1}^{Z}(\sum_{u \in \mathcal{N}(v)}\alpha_{uv}^{z} \cdot \mathbf{W}^{z} \cdot \mathbf{h}_{u}^{l-1}) \label{eq:gat}
\end{equation}
where $\Vert$ is the concatenation operation, $Z$ is the total number of projection heads in the attention mechanism, $W^z$ is the linear transformation weight matrix, and $\alpha_{uv}^{z}$ is the attention coefficient calculated by the $z$-th projection head.
\item  \textit{Graph Isomorphism Network (GIN).}
GraphSAGE can be treated as an instance of the Weisfeiler-Lehman test.
Xu~et al.~\cite{XHLJ19} propose Graph Isomorphism Network (GIN) to extend GraphSAGE with arbitrary aggregation functions on
multi-sets. 
GIN is theoretically proven to be as powerful as the Weisfeiler-Lehman test of graph isomorphism.
Its aggregation function can be represented as:
\begin{equation}
\mathbf{h}_{v}^{l} = (1+\epsilon^{l-1}) \cdot \mathbf{h}_{v}^{l-1}+\sum_{u \in \mathcal{N}(v)} \mathbf{h}_{u}^{l-1} \label{eq:gin}
\end{equation}
where $\epsilon$ is a learnable parameter to adjust the weight of node $v$.
\end{itemize}

The inductive GNNs usually employ shared weight parameters and neighborhood sampling to speed up the computation.

\mypara{Responses by Inductive GNNs}
An inductive GNN model learns the parameters of aggregation functions in different layers from the training data.
Once trained, the learned GNN model can infer previously unseen data. 
Such capability paves the way for remotely deployed GNN models in the wild (e.g., GROVER~\cite{RBXXWHH20}, DGL~\cite{WZYGLSZMYGXHKLZ19}) to make inferences on graphs for customers via publicly accessible API.
Moreover, a trained GNN model is often used to perform node embedding tasks, and the resulted node embeddings can then help to perform other downstream ML tasks (e.g.,  fine-tuning pretrained GNNs~\cite{HLGZLPL20}, model partitioning~\cite{SS19}) or graph visualization.
Therefore, in this paper, we consider three query responses from a target GNN model when facing a query node, namely predicted posterior probability,  embedding vector, and 2-dimensional t-SNE projections from the embedding~\cite{GF18,ZCZYLS18,LRKAK19,ZCZ20,WPCLZY20}.
Specifically, given a query node $v \in \mathbf{V}_Q$, we feed its $l$-hop subgraph (i.e., $G_v^l$) to a remote GNN model and obtain one of the three corresponding responses. 
All the query nodes with the edges between them can form a query graph/dataset, denoted as $\mathbf{G}_Q$.
Note that $\mathbf{G}_Q$ is not necessarily connected and for each query node, the $l$-hop subgraph is extracted from $\mathbf{G}_Q$ only.

% ----------------------------------------------------
\section{Threat Model}
\label{sec:threat_model}
% ----------------------------------------------------

In this section, we outline the threat model to characterize the adversary's background knowledge and the goal of the model stealing attack.

% ----------------------------------------------------
\subsection{Attack Setting}
% ----------------------------------------------------

We frame our attack in a \textit{black-box} setting, which is the most challenging scenario for the adversary mentioned in previous work~\cite{SSSS17,OASF18,JCBKP20,HJBGZ21,HWWBSZ21}.
That is, the adversary has no knowledge of the target GNN model (e.g., model parameters, model architecture) and cannot tamper with its training process (e.g., training graph $\mathbf{G}_O$).
Our attack setting is fundamentally different from the previous attacks~\cite{DR19,WYPY20}.
These attacks assume that the adversary can gain access to the target model’s training process.
However, such a strong assumption is unrealistic in the real world as it is impractical to expect an adversary can interfere with the target model at its training time.
Note that in this paper we focus on the aforementioned three node-level query responses. 
The target model is an inductive GNN model which accepts node $v$'s $l$-hop subgraph $\mathbf{G}_v^l$ as input and returns the corresponding response for the given node $v$, i.e., its predicted posterior probability, node embedding vector, or 2-dimensional t-SNE projection.

% ----------------------------------------------------
\subsection{Adversary's Goal}
% ----------------------------------------------------

Following the taxonomy defined by Jagielski~et al.~\cite{JCBKP20}, the adversary's goal falls into two categories, i.e., theft and reconnaissance. 

\begin{itemize}
\item The goal of the theft adversary is to build a surrogate model $\mathcal{M}_S$ that matches the accuracy of the target model $\mathcal{M}_T$ on the target task~\cite{TZJRR16,PMGJCS17}.
The theft adversary's motivation is compromising the intellectual property and violating the confidentiality of the target model $\mathcal{M}_T$.
\item Subtly different from the theft adversary, the reconnaissance adversary aims to build a surrogate model that closely matches the behavior of the target model.
That is, $\mathcal{M}_S$ seeks an agreement to $\mathcal{M}_T$ on any input.
A surrogate model $\mathcal{M}_S$ with high fidelity to $\mathcal{M}_T$ enables the adversary to leverage it as a stepping stone to launch further attacks.
For instance, the adversary can craft adversarial examples using this $\mathcal{M}_S$ instead of risking potentially detectable queries to the target model  $\mathcal{M}_T$~\cite{PMGJCS17}.
\end{itemize}

Note that the reconnaissance adversary's motivation is faithfully copying the behavior of $\mathcal{M}_T$ (e.g., $\mathcal{M}_S$ and $\mathcal{M}_T$ may make the same wrong/correct prediction of an input) though both adversaries intend to get close to the performance (i.e., accuracy) of the target model. 
We refer the audience to the work by Jagielski~et al.~\cite{JCBKP20} for additional discussions.

% ----------------------------------------------------
\subsection{Adversary's Capability}
\label{sec:adversary_capability}
% ----------------------------------------------------

We first assume an adversary can make queries to a target model $\mathcal{M}_T$.
Given a query graph $\mathbf{G}_Q$, the adversary can query all its nodes with their corresponding $l$-hop subgraphs, then obtain the query responses $\mathbf{R}$.
This assumption is in line with the adversarial machine learning setting, whereas the attackers exploit the remote target model $\mathcal{M}_T$ via publicly accessible API~\cite{TZJRR16,OSF19,HJBGZ21,HWWBSZ21}. 
The response $\mathbf{R}$ of the whole query graph $\mathbf{G}_Q$, depending on the target model's specification, may be returned in the form of a node embedding matrix (denoted as $\mathbf{H}$), a predicted posterior probability matrix (denoted as $\boldsymbol{\Theta}$), or a t-SNE projection matrix of $\mathbf{H}$ (denoted as $\boldsymbol{\Upsilon}$), where each row is a 2-dimensional vector. 
These three responses are representative of the real-world API output.
For example, t-SNE and node embeddings are widely returned in the scenarios of graph visualization~\cite{HZYGH20}, transfer learning~\cite{ZXWZHY20}, federated learning~\cite{HBCRZHAA21}, fine-tuning pretrained GNNs~\cite{HLGZLPL20}, and model partitioning where the target model is split into local and cloud parts bridged by embeddings information~\cite{SS19}.
Moreover, these responses characterize three different levels of knowledge the adversary may gain access to in practice.
For instance, the t-SNE projection matrix $\boldsymbol{\Upsilon}$ is usually used for data visualization. 
It contains the least information that an adversary can harvest.
We later show that the adversary can still launch the attack through such data visualization function of a remote model. 

Second, we assume that the query graph $\mathbf{G}_Q$ (both node features $\mathbf{X}_Q$ and graph structure $\mathbf{A}_Q$) is from the same distribution of the training graph $\mathbf{G}_O$ used to train the target model $\mathcal{M}_T$. 
Here we consider the term ``same distribution'' as $\mathbf{G}_O$ and $\mathbf{G}_Q$ are drowned randomly from the same dataset. 
We do not necessarily require $\mathbf{G}_O$ and $\mathbf{G}_Q$ to have the same graph characteristics. 
Besides, the nodes in the query graph $\mathbf{G}_Q$ does not need to be in the training graph $\mathbf{G}_O$.
For example, both $\mathbf{G}_O$ and $\mathbf{G}_Q$ can be subgraphs sampled from social networks like Twitter but there is no overlap with each other.
This assumption is in line with recent attacks to neural networks where an adversary uses part of a public dataset to exploit the target model~\cite{JCBKP20,HJBGZ21} and graphs are available in some domains (e.g., social networks and molecular graphs). 
We further relax this assumption and demonstrate that the adversary can still launch effective attacks even without the graph structural information of $\mathbf{G}_Q$. 
This weak assumption of adversary knowledge makes our attack more practical in real-world scenarios.
For instance, the adversary may compromise a user database and acquire their profile information.
However, the relationships among the users, i.e., the graph structure, may not be revealed.
We later demonstrate that the adversary can still launch high-performance model stealing attacks in this scenario (see \autoref{sec:evaluation}).

\noindent \textbf{Notes}. Our threat model is different from the \emph{causative} and \emph{evasion} adversarial attacks to GNNs~\cite{SDYWYHL18,DLTHWZS18,CLPXCXHZ20,JLXWT20,XMLDLTJ20,XPJW21}. 
Those attacks allow the attackers to manipulate the training graphs in order to change the parameters of the target model, or modify the node features and/or graph structure to fool the inductive GNN models.
Our attack is an instance of \emph{exploratory attacks}.
We do not tamper with the original training process.
The goal is thereof not to change the parameters or fool the target GNN model. Instead, it is designed to steal a copy of a target inductive GNN model.

Our attack also differs from the existing model stealing attacks against \emph{transductive GNNs}~\cite{DR19,WYPY20} in several key aspects. 
First, these attacks~\cite{DR19,WYPY20} focus on attacking transductive GNN models.
Both methods assume that the query graph is part of the graph used for training the target GNN model and must be involved in the training process, hence unrealistic. 
Our attacks instead focus on a more realistic stealing attack scenario whereas the adversary only queries the target model via remotely accessible API.
We do not tamper with the training process of the target model.
In turn, our threat model is practical and fills the gap to understand if both theft and reconnaissance adversaries can steal inductive GNNs with high accuracy and high fidelity.
Moreover, the existing methods~\cite{DR19,WYPY20} are limited to the GCN model (i.e., model dependent) and rely on node predicted posterior probability scores to launch attacks.
In contrast, our attack is model agnostic and can still successfully copy the target model’s behavior with marginal information.
Besides, our attack setting is different from Attack-3 proposed by Wu~et al.~\cite{WYPY20}.
Specifically, Attack-3 in~\cite{WYPY20} trains the surrogate model without querying the target model, while our attacks do interact with the target model.
Note that it is impractical to attain the reconnaissance goal without interacting with the target model~\cite{LZ21,HWWBSZ21}. 
Also, the target model considered by Wu~et al.~\cite{WYPY20} is a transductive GNN while ours focuses on the inductive GNN (the difference between transductive and inductive GNN is described in \autoref{sec:background}).

% ----------------------------------------------------
\section{Model Stealing Attack}
\label{sec:model_stealing_attack}
% ----------------------------------------------------

In this section, we first outline 6 attack scenarios that can be launched by the adversary given different levels of knowledge.
Then we propose our attack framework.

% ----------------------------------------------------
\subsection{Attack Taxonomy}
\label{sec:attack_taxonomy}
% ----------------------------------------------------

\begin{figure}[!t]
\centering
\includegraphics[width=0.8\linewidth]{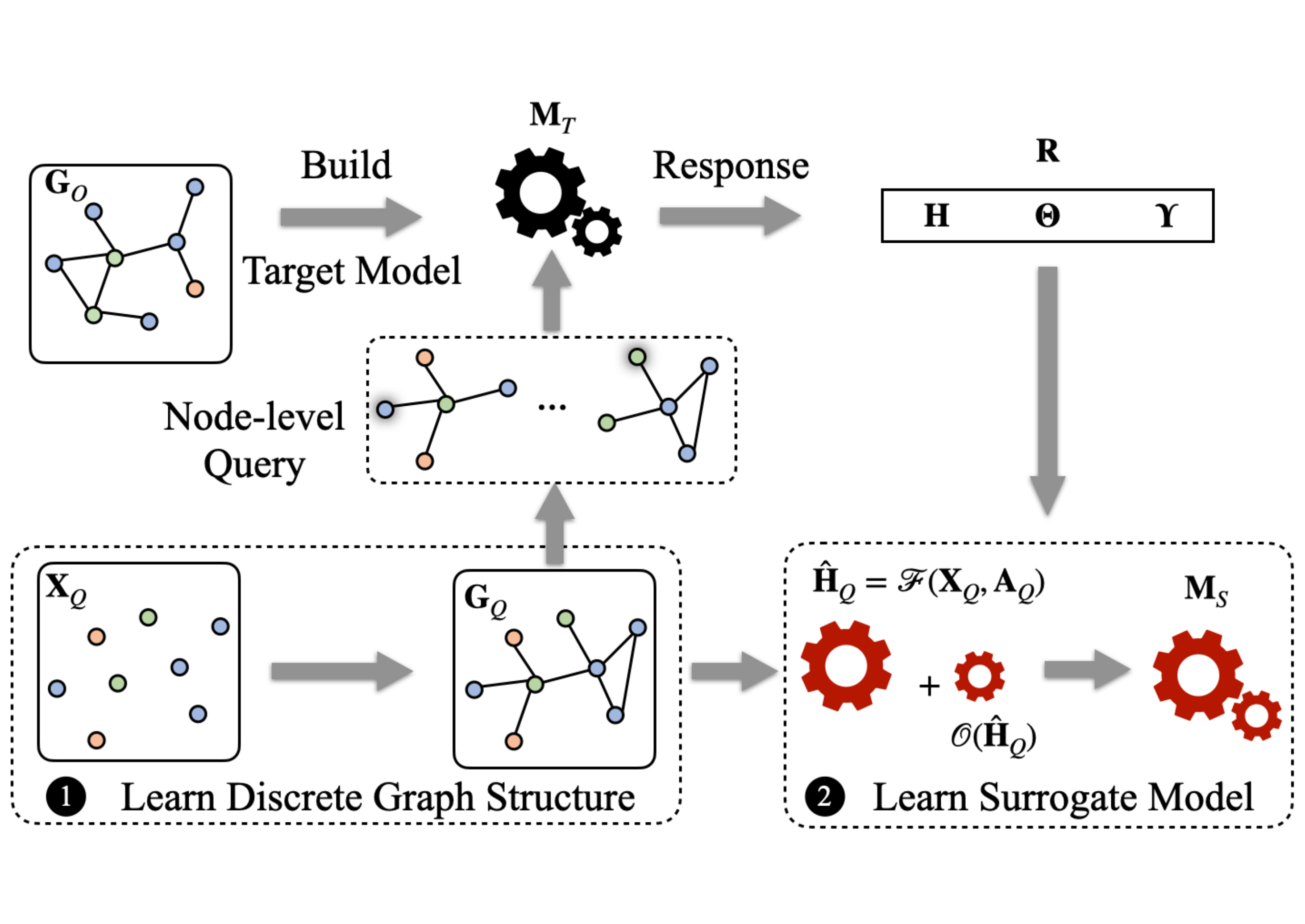} 
\caption{Overview of model stealing attack against inductive GNNs.}
\label{fig:surrogate_model}
\end{figure}

As outlined in \autoref{sec:threat_model}, the adversary has two main pieces of information at their disposal to launch the model stealing attack: the query graph $\mathbf{G}_Q=(\mathbf{A}_Q, \mathbf{X}_Q, \mathbf{C}_Q)$ and its corresponding query response $\mathbf{R}$ (i.e., node embedding matrix $\mathbf{H}$, predicted posterior probability matrix $\boldsymbol{\Theta}$, or t-SNE projection matrix $\boldsymbol{\Upsilon}$).
Recall that the adversary may not have graph structural information of the query graph $\mathbf{G}_Q$ (i.e., the adjacency matrix $\mathbf{A}_Q$ may be missing). 
We thus have 6 possible attack scenarios falling into two types (see \autoref{tab:attack_taxonomy}).

\begin{itemize}
\item \textbf{Type I Attack}. 
When the adversary obtains the query graphs which are of the same distribution as the training graphs, they can launch a Type I attack against the target model.
The application scenarios of Type I attack include side-effects prediction due to drug interactions~\cite{ZAL18}, financial fraud detection~\cite{LAQCFYH21}, recommendation systems~\cite{WYCLHW20}, etc.
In these cases, the adversary can obtain the query graphs (e.g., drug-protein interaction graphs, transaction graphs, user-item purchase graphs) that are of the same distribution of the training graphs used to train the target models due to the wide availability of such graphs.
\item \textbf{Type II Attack.}
When the graph structural information is missing, the adversary can resort to a Type II attack to steal the target model.
For instance, social networks such as Instagram or Tinder do not reveal the social relationship of (private) user accounts.
However, an adversary can crawl users' profile information without social relations (i.e., graph structural information).
Then they leverage Type II attacks to rebuild a relationship graph and steal the target models offered by these service providers.
Besides, Type II attacks can also be used in stealthy insider threat scenarios.
For instance, a company may enforce data segregation due to data privacy and security concerns (e.g., one department has user information and another has the relationship graph). 
To train a joint model, the company needs to perform vertical federated learning~\cite{YLCT19}.  
The insider attackers may leverage the information they can access (e.g., user information) to steal the joint model using our Type II attack.
\end{itemize}

It is straightforward to observe in \autoref{tab:attack_taxonomy} that the attacks are increasingly harder in the row order.
For instance, the Type II.3 attack is the most challenging scenario from the adversary's perspective since they only have a set of node features to start with.
As such, they first need to restore the relationships among the nodes and build the query graph $\mathbf{G}_Q$, then leverage the t-SNE projection matrix obtained from $\mathcal{M}_T$ to steal the target model.
Besides, model stealing attacks against the classical ML models focus on the scenario that the remote models return the predicted posterior probability.
We note that the node embedding and the t-SNE projection-based query responses are the new attack surface of the inductive GNN models and outline the technical details in \autoref{sec:framework}. 

% ----------------------------------------------------
\subsection{Attack Framework}
\label{sec:framework}
% ----------------------------------------------------

To tackle the aforementioned challenges, we propose a unified attack framework as illustrated in \autoref{fig:surrogate_model} to launch both Type I and II attacks.
It has two components.
The first component (\ding{182} in \autoref{fig:surrogate_model}) learns the missing graph structure $\mathbf{A}_Q$.
The output of the first component is a learned query graph $\mathbf{G}_Q$.
This component is designed to facilitate Type II attacks, hence not required for Type I attacks.
It is important to note that this reconstruction process is done locally at the attacker's side. 
They do not interact with the target model.
The second component (\ding{183} in \autoref{fig:surrogate_model}) learns a surrogate model from the response of the target model given the query graph $\mathbf{G}_Q$.  
The output of the second component is a learned surrogate model $\mathcal{M}_S$.
We outline their details below.

\begin{table}[t]
\centering
\caption{Attack Taxonomy. 
The attacks are increasingly harder in row order.}
\scalebox{0.9}{
\begin{tabular}{ccccc}
\toprule
\multirow{2}{*}{\textbf{Attack}} & \multirow{2}{*}{$\mathbf{A}_Q$}    & \multicolumn{3}{c}{Response $\mathbf{R}$} \\ \cmidrule(lr){3-5}
            &        & $\mathbf{H}$ & $\boldsymbol{\Theta}$  & $\boldsymbol{\Upsilon}$       \\
\midrule
\textbf{I.1}  & \ding{51}        & \ding{51}       &           &       \\
\textbf{I.2}  &  \ding{51}        &            & \ding{51} &           \\
\textbf{I.3}  & \ding{51}       &           &           &  \ding{51}  \\
\midrule
\midrule
\textbf{II.1} & \ding{55}     &       \ding{51} &           &           \\
\textbf{II.2} &\ding{55}        &    & \ding{51} &           \\
\textbf{II.3} & \ding{55}     &     &           &  \ding{51} \\
\bottomrule
\end{tabular}
}
\label{tab:attack_taxonomy}
\end{table}

% ----------------------------------------------------
\subsubsection{Learn Discrete Graph Structure (\ding{182})} 
% ----------------------------------------------------

Recall that the adversary does not have the adjacency matrix $\mathbf{A}_Q$ for the query graph $\mathbf{G}_Q$ at their disposal in Type II attacks (see \autoref{sec:threat_model}).
However, the graph structure is necessary for the adversary to conduct the attack.
As such, without a graph structure to piece $\mathbf{X}_Q$ together, the adversary must first build the adjacency matrix $\mathbf{A}_Q$ from query graph $\mathbf{G}_Q$ and then query the target model $\mathcal{M}_T$.
The goal of this component is thereof to learn a high-quality discrete graph structure that enables the adversary to query and gain useful knowledge from the query response returned by the target model.

One common approach is to create a $k$-nearest neighbor ($k$NN) graph from node features $\mathbf{X}_Q$.
However, the efficacy of the resulting $k$NN graph and consequent GNN model rests on the choice of $k$ (e.g., $k$NN graphs often produce nodes with extremely high degrees) and the chosen similarity measure over the node features.
To this end, we leverage the IDGL framework proposed by Chen~et al.~\cite{CWZ20} to learn a query graph $\mathbf{G}_Q$ by minimizing a joint loss function combining both the task-dependent prediction loss and the graph regularization loss.
Note that task-dependent prediction loss is flexible, and can be tailored to use different loss functions (e.g., node classification loss or link prediction loss).
The graph regularization loss controls the smoothness, connectivity, and sparsity of the resulting graph.
Additional details can be found in~\cite{CWZ20}.

To launch Type II attacks, the adversary first initiates a $k$NN graph constructed based on multi-head weighted cosine similarity, then utilizes the IDGL framework to search for a hidden graph structure that augments the initial $k$NN graph structure using the aforementioned joint loss function.
Note that the $k$NN graph is only used to seed the initial graph structure.
The final learned graph structure is optimized during the learning process and may not contain the edges from the original $k$NN graph.
This component's output is a learned query graph $\mathbf{G}_Q$.

% ----------------------------------------------------
\subsubsection{Learn Surrogate Model (\ding{183})}
% ----------------------------------------------------

The goal of the second component is to learn a surrogate GNN model from the query response returned by the target model given a query graph $\mathbf{G}_Q$.  
To this end, we first discuss our observation of three state-of-the-art inductive GNN models, namely \textit{GraphSAGE}, \textit{GIN}, and \textit{GAT}.
We then propose a unified framework to learn surrogate models given both Type I and Type II attacks.

\mypara{Observation} 
According to the convolution operations defined on graphs, graph neural networks can be categorized as spectral approaches and spatial approaches~\cite{WPCLZY20,CCZJFZCL20,ZWSJC21,ZCZYLS18}.
For spectral approaches, the graph is represented with a Laplacian matrix according to the spectral theories with the convolution operation defined in the sequence domain via Fourier transform.
Different from spectral approaches, spatial approaches define graph convolutions by collective information propagation (i.e., propagating node information along edges) and perform convolution by considering node neighborhoods.
Leveraging the insights from~\cite{WPCLZY20,CCZJFZCL20,ZWSJC21,ZCZYLS18}, in spatial-based GNN methods, at $l$-th layer, node embedding $\mathbf{h}_v^{l}$ is iteratively updated using \autoref{eq:iterative_update} where $\Phi$ and $\Psi$ are weight functions, and $\eta$ is a normalization factor. 

\begin{equation}
\mathbf{h}_v^{l} = \Phi(v) \mathbf{h}_{v}^{l-1} +  \sum_{u \in \mathcal{N}(v)} \Psi(u) \frac{\mathbf{h}_u^{l-1}}{\eta} \label{eq:iterative_update}
\end{equation}
If $\Psi$ assigns the same weight to all neighbors \autoref{eq:iterative_update} can be rewritten in matrix form as \autoref{eq:iterative_matrix}, where $\mathbf{I}$ is an identity matrix and $\tilde{\mathbf{A}}$ is a normalized form of adjacency matrix $\mathbf{A}$.
Note that $\mathbf{H}^{0}=\mathbf{X}$.

\begin{equation}
\mathbf{H}^{l} = (\Phi \mathbf{I} + \Psi \tilde{\mathbf{A}}) \mathbf{H}^{l-1} \label{eq:iterative_matrix}
\end{equation}
It is straightforward to prove that GraphSAGE (\autoref{eq:graphsage} in \autoref{sec:background}) can be written in matrix form as \autoref{eq:graphsage_matrix}, where   $\Phi=1$, $\Psi=1$, and $\tilde{\mathbf{A}}$ is a random-walk normalized Laplacian matrix. 

\begin{equation}
\mathbf{H}^{l} = (\mathbf{I} + \tilde{\mathbf{A}}) \mathbf{H}^{l-1} \label{eq:graphsage_matrix}
\end{equation}
Similarly, GIN (\autoref{eq:gin} in \autoref{sec:background})  can be written as \autoref{eq:gin_matrix}, where $\Phi=1+\epsilon$, $\Psi=1$, $\mathbf{A}$ is unnormalized adjacency matrix.

\begin{equation}
\mathbf{H}^{l} = ((1+\epsilon) \mathbf{I} + \mathbf{A} ) \mathbf{H}^{l-1} \label{eq:gin_matrix}
\end{equation}
In the same way, GAT (\autoref{eq:gat} in \autoref{sec:background}) can be written as \autoref{eq:gat_matrix}, where $\Psi=\boldsymbol{\Xi}$ is a learnable neighbor weight matrix and $\Phi=0$.

\begin{equation}
\mathbf{H}^{l} = (\boldsymbol{\Xi} \otimes \mathbf{A} ) \mathbf{H}^{l-1} \label{eq:gat_matrix}
\end{equation}
Given \autoref{eq:graphsage_matrix}, \autoref{eq:gin_matrix}, and \autoref{eq:gat_matrix}, we can see that there exists an intrinsic connection among GraphSAGE, GAT, and GIN (i.e., they are special cases of \autoref{eq:iterative_matrix}). 
It is evidential that these models are spatial-based GNN models through adopting different designs for feature aggregation.  
This observation enables us to design a unified attack framework to leverage the query graph $\mathbf{G}_Q$ as the data and the query response $\mathbf{R}$ as the supervised information to build the surrogate model in a unified manner.

\mypara{Learning Framework}
Our attack framework is illustrated in \autoref{fig:surrogate_model}. 
The surrogate model $\mathcal{M}_s$ in \ding{183} consists of two modules. 
The first module is a customized inductive GNN model (denoted as $\mathcal{F}$) taking all nodes' $l$-hop subgraphs from $\mathbf{G}_Q$ as the training data and the query response $\mathbf{R}$ as the supervised information.
Since the responses from $\mathcal{M}_T$ are vectors in Euclidean space, they reflect the spatial connectivity among nodes either from the graph connectivity perspective (t-SNE projection or node embedding), or from the node label perspective (the predicted posterior probability). 
That is, the nodes that are close or connected in the query graph $\mathbf{G}_Q$ should be close in $\boldsymbol{\Upsilon}$ or $\mathbf{H}$, and the nodes that are of the same label should be close in $\boldsymbol{\Theta}$.

Following the above observation, the key idea of our attack is that we ignore the response types and uniformly treat all three possible responses as embedding vectors.
As such, for the first module, the goal is to minimize the RMSE loss ($\mathcal{L}_R$) between $\hat{\mathbf{H}}_Q$ and $\mathbf{R}$ as shown in \autoref{eqn:component_1}.

\begin{eqnarray}
\hat{\mathbf{H}}_Q & = &  \mathcal{F} (\mathbf{X}_Q, \mathbf{A}_Q)   \nonumber \\
\mathcal{L}_R  & = & \frac{1}{n_Q} \| \hat{\mathbf{H}}_{Q} - \mathbf{R} \|_{2,1} \label{eqn:component_1}
\end{eqnarray}
where $n_Q$ denotes the number of nodes of the query graph $G_Q$.
Here $\hat{\mathbf{H}}$ keeps the same dimension as $\mathbf{R}$.
The rationale of using the RMSE loss is that $\mathcal{F}$ is optimized to maintain the similar spatial connectivity among the nodes in $\hat{\mathbf{H}}_Q$ as suggested by $\mathbf{R}$.
Note that the output from the first component cannot be directly used for node classification tasks.
In light of this, we employ an MLP as the classifier (denoted as $\mathcal{O}$). 
It takes the output from the first module (i.e., $\hat{\mathbf{H}}_Q$) as the input and $\mathbf{C}_Q$ as the supervision information to minimize the prediction error ($\mathcal{L}_P$) as shown in \autoref{eqn:component_2}. 

\begin{equation}
\mathcal{L}_P =  - \frac{1}{n_Q} \underset{ {v \in \mathbf{G}_Q} } \sum  \underset{ {i \in |\mathbf{C}_Q|} } \sum   c_i log [ \mathcal{O}(\mathbf{h}_v)_i ]   \label{eqn:component_2}
\end{equation}

We optimize the first module with $\mathcal{L}_R$, then we freeze it and optimize the second module with $\mathcal{L}_P$.
The two modules are then chained together as our surrogate model.
Followed by Orekondy~et al.~\cite{OSF19}, we assume that $\mathbf{C}_Q$ is obtained by the adversary.

Our attack design enjoys two-fold flexibility. 
\emph{First}, our attack allows the adversary to conduct the attack without knowing the target model's architecture. 
This makes the threat model closer to the real-world scenario where the adversary only has query access to the target model.
\emph{Second}, our attack framework shows that node embedding and 2-dimensional t-SNE projection as the query response can be the new attack surface of the inductive GNN models.
Notably, t-SNE projection is for data visualization purposes.
It barely unveils the graph structural information to the adversary.
However, we demonstrate that our attack can still successfully copy the target model's behavior from such marginal information.

% ----------------------------------------------------
\section{Evaluation}
\label{sec:evaluation}
% ----------------------------------------------------

In this section, we perform an in-depth analysis of the proposed model stealing attack against inductive GNNs. 
We first introduce the experimental setup, and present the evaluation results for Type I and Type II attacks from both theft and reconnaissance adversary's perspectives.
We then explore how various query budgets may affect the attack performance.
Finally, we study how different hyperparameters of the surrogate model may influence the attack performance.

% ----------------------------------------------------
\subsection{Experimental Setup}
\label{sec:experimental_setup}
% ----------------------------------------------------

\begin{table}[!t]
\centering
\caption{Summary of datasets.}
\scalebox{0.9}{
\begin{tabular}{lccccc}
\toprule
\textbf{Dataset}                                                                      & $|\mathbf{V}|$ & $|\mathbf{E}|$ & $|\mathbf{X}|$ & $|\mathbf{C}|$ & \textbf{Density} \\
\midrule
\textbf{DBLP}   & 17,716 & 105,734 & 1,639 & 4 & 0.0007         \\
\textbf{Pubmed}  & 19,717 & 88,648 & 500 & 3 & 0.0005         \\
\textbf{Citeseer}   & 4,230 & 5,358 & 602 & 6 & 0.0006 \\
\textbf{Coauthor}    & 34,493 & 495,924 & 8,415 & 5 & 0.0008         \\
\textbf{ACM}    & 3,025 & 26,256 & 1,870 & 3 & 0.0057         \\
\textbf{Amazon}    & 7,650 & 143,663 & 745 & 8 & 0.0049       \\
\bottomrule
\end{tabular}
}
\label{tab:graph_stats}
\end{table}

\mypara{Datasets} 
We use 6 public datasets to evaluate the performance of our attack, including DBLP~\cite{PWZZW16}, Pubmed~\cite{SNBGGE08}, Citeseer Full (abbreviated as Citeseer)~\cite{GBL98}, Coauthor Physics (abbreviated as Coauthor)~\cite{SMBG18}, ACM~\cite{WJSWYCY19}, and Amazon Co-purchase Network for Photos (abbreviated as Amazon)~\cite{MTSH15}.
These datasets are widely employed as benchmark datasets to evaluate the performance of GNNs~\cite{KW17,HYL17,XHLJ19}.
Among them, DBLP, Pubmed, and Citeseer are citation networks with nodes representing publications and edges indicating citations among these publications.
Coauthor is a user interaction network where nodes represent the users and edges represent interactions between them.
ACM and Amazon are cooperative networks where nodes represent the papers/items and there is an edge between two nodes if they have the same author or purchased together.
We use these datasets to verify the efficacy of our attacks given different graph characteristics (e.g., graph size, node feature size, number of classes, etc.). 
Statistics of these datasets are summarized in \autoref{tab:graph_stats}. 

\mypara{Dataset Configuration}
For each dataset, we split them into three parts.
The first part consists of 20\% randomly sampled nodes that are used to train the target model $\mathcal{M}_T$.
The second part consists of 30\% randomly sampled nodes, forming our query graph $\mathbf{G}_Q$.
We further show that our attacks are still effective with fewer nodes to form the query graph (see \autoref{fig:pubmed_query_budget}).
The third part consists of the rest 50\% of the nodes, functioning as the testing data for both $\mathcal{M}_T$ and $\mathcal{M}_S$.
This setting matches the inductive learning on evolving graphs as laid out in~\cite{HYL17}.

\mypara{Target Model ($\mathcal{M}_T$)}
We use GIN, GAT, and GraphSAGE as our target models' architectures in our evaluation.
For reproducibility purposes, we outline the details below.

\begin{itemize}
\item \textbf{GIN.}
We use a 3-layer GIN model with a fixed neighborhood sample size of 10 at each layer.
For the first hidden layer, we set the hidden unit size to 256.
For the second layer, we set the hidden unit size to the embedding size (i.e., 64, 128, or 256 in our experimental setting).
The final layer is used for classification.
\item \textbf{GAT.}
We use a 3-layer GAT model with a fixed neighborhood sample size of 10 at each layer. 
The first layer consist of 4 attention heads and the hidden unit size is 256.
The second layer consists of 4 attention heads and the hidden unit size is the embedding size.
The final layer is used for classification following the original design~\cite{VCCRLB18}.
\item \textbf{GraphSAGE.} 
Following Hamilton~et al.~\cite{HYL17}, we use a 2-layer GraphSAGE with neighborhood sample sizes of 25 and 10 respectively.
For the first hidden layer, we set the hidden unit size to 256.
For the second layer, we set the hidden unit size to the embedding size.
Each layer employs a GCN aggregator and uses 0.5 dropout rate to prevent overfitting.
Finally, we use a linear transformation layer for classification.
\end{itemize}

All models use cross-entropy as the loss function, ReLU as the activation function between layers, and Adam as the optimizer with an initial learning rate of 0.001.
We train all models for 200 epochs and select the best models with the highest validation accuracy.
All models above follow the design specifications outlined in the respective papers.

\mypara{Query Response ($\mathbf{R}$)} 
For the node embedding $\mathbf{H}$, we fix the sizes to three commonly used values, i.e., 64, 128, and 256.
For the node predicted posterior probability $\boldsymbol{\Theta}$, the dimension sizes (i.e., the number of classes) are dataset dependent and outlined in \autoref{tab:graph_stats}. 
For the node projection $\boldsymbol{\Upsilon}$, we use t-SNE to project the node embedding $\mathbf{H}$ into 2-dimensional vectors, which cover most of the data visualization cases. 

\mypara{Surrogate Model ($\mathcal{M}_S$)}
Recall our attack design in \autoref{sec:attack_taxonomy}, the surrogate model consists of two components to provide extensibility.
The first component is a customized GNN model taking the subgraph extracted from $\mathbf{G}_Q$ as the input and using the RMSE as its loss function. 
The second component is a 2-layer MLP with the hidden unit size of 100. 
It takes the output from the first component as the input and uses the cross-entropy as its loss function.
Both components use Adam optimizer with a learning rate of 0.001.
We train the first and the second components for 200 epochs and 300 epochs, respectively.
For evaluation purposes, we use customized GraphSAGE, GAT, and GIN models as the first component in our surrogate model.
The details are outlined below.

\begin{itemize}
\item \textbf{GIN.}
We use a 2-layer GIN model with neighborhood sample sizes of 10 and 50 respectively. 
The hidden unit size is 256 for the first layer.
For the second layer, we set the hidden unit size to the size of the query response.
\item \textbf{GAT.}
We use a 2-layer GAT model with neighborhood sample sizes of 10 and 50 respectively. 
Both the first and the second layers consist of 4 attention heads and we follow the same hidden unit size as GIN we mentioned above.
\item \textbf{GraphSAGE.} 
We use a 2-layer GraphSAGE with neighborhood sample sizes of 10 and 50 respectively and follow the same hidden unit size as GIN we mentioned above.
\end{itemize}

\mypara{Query Graph Reconstruction Configuration}
We use the IDGL framework proposed by Chen~et al.~\cite{CWZ20} to learn the missing graph structural information, i.e., $\mathbf{A}_Q$, of the query graph $\mathbf{G}_Q$ for Type II attacks.
For simplicity, we use node classification loss and leave experimenting with other loss functions as future work.
The initial $k$ of $k$NN graph is set to 24.
We use a 2-layer (hidden unit size is set to 256) GraphSAGE with GCN aggregator to learn the node embedding.
Weighted cosine similarity matrices with 8 attention heads are employed to decide if there exists an edge between two node embeddings during the graph learning process.
The graph structure and the GraphSAGE parameters are jointly and iteratively learned by minimizing a hybrid loss function combining both the node classification loss and the graph regularization loss.
We set the cutoff value $\epsilon$ to 0.99 to identify the final edges in the learned adjacency matrix $\mathbf{A_Q}$ of the query graph $\mathbf{G}_Q$. 

\mypara{Metrics}
Following the taxonomy defined by Jagielski~et al.~\cite{JCBKP20}, we use two metrics, i.e., accuracy and fidelity, to evaluate the performance of our attack.
We use accuracy (i.e., the number of correct predictions made divided by the total number of predictions made) as our evaluation metric of theft adversary.
Accuracy has been dominantly used in evaluating node classification performance of GNNs~\cite{KW17,HYL17,VCCRLB18}.
Recall that the goal of reconnaissance adversary is to closely match the behavior of the target model (see \autoref{sec:threat_model}), we use fidelity (i.e., the number of predictions agreed by both $\mathcal{M}_S$ and $\mathcal{M}_T$) as the second evaluation metric of our attack~\cite{JSMA19,JCBKP20}.
Both metrics are normalized between 0 and 1.
Higher scores imply better performance.

\mypara{Runtime Configuration}
Note that we have 27 different combinations for each response in each dataset (i.e., combinations of three target models, three responses, and three surrogate models).
All the experiments in this paper are repeated 5 times.
For each run, we follow the same data configuration and report the mean as well as the standard deviation of the aforementioned two metrics to evaluate the attack performance.

% ----------------------------------------------------
\subsection{Performance Evaluation: Type I Attacks}
\label{sec:type1_performance}
% ----------------------------------------------------

\begin{table}[t]
\centering
\caption{The performance of the original classification tasks on all the 6 datasets using 3 different GNN structures.}
\scalebox{0.9}{
\begin{tabular}{lccc}
\toprule
\multirow{2}{*}{\textbf{Dataset}} & \multicolumn{3}{c}{$\mathcal{M}_T$} \\
\cmidrule{2-4}
                         & \textbf{GIN}       & \textbf{GAT}       & \textbf{SAGE}       \\
\midrule
\textbf{DBLP}  &  0.872 & 0.838 & 0.858 \\
\textbf{Pubmed}  &  0.924 & 0.905 & 0.909 \\
\textbf{Citeseer}  &  0.910 & 0.910 & 0.918 \\
\textbf{Coauthor}  &  0.953 & 0.965 & 0.956 \\
\textbf{ACM}  &  0.929 & 0.935 & 0.937 \\
\textbf{Amazon}  &  0.856 & 0.953 & 0.937 \\
\bottomrule
\end{tabular}
}
\label{tab:acc_target_models}
\end{table}

\begin{table*}[t]
\centering
\caption{The accuracy and fidelity scores of Type I attacks using different response information on all the 6 datasets. 
Both average values and standard deviations are reported. 
The accuracy differences (in parenthesis) of the surrogate models to the target models are also reported. 
We use GraphSAGE as the surrogate model. }
\scalebox{0.8}{
\begin{tabular}{lccccccc}
\toprule
\multirow{3}{*}{\textbf{Dataset}} &
  \multirow{3}{*}{\begin{tabular}[x]{@{}c@{}}$\mathcal{M}_S$\\(SAGE)\end{tabular}} &
  \multicolumn{6}{c}{$\mathcal{M}_T$} \\
\cmidrule{3-8}
 &
   &
  \multicolumn{2}{c}{\textbf{GIN}} &
  \multicolumn{2}{c}{\textbf{GAT}} &
  \multicolumn{2}{c}{\textbf{SAGE}} \\
 \cmidrule(lr){3-4} \cmidrule(lr){5-6} \cmidrule(lr){7-8}
 &
   &
  \textbf{Accuracy} &
  \textbf{Fidelity} &
  \textbf{Accuracy} &
  \textbf{Fidelity} &
  \textbf{Accuracy} &
  \textbf{Fidelity} \\

\midrule
              &  Projection  &  0.704$\pm$0.030 (-0.168) & 0.727$\pm$0.032 & 0.682$\pm$0.003 (-0.156) & 0.690$\pm$0.002 & 0.708$\pm$0.018 (-0.150) & 0.748$\pm$0.018 \\
              &  Prediction  &  0.769$\pm$0.006 (-0.103) & 0.799$\pm$0.006 & 0.787$\pm$0.006 (-0.051) & 0.827$\pm$0.005 & 0.810$\pm$0.005 (-0.048) & 0.884$\pm$0.004 \\
\multirow{-3}{*}{\textbf{DBLP}}  &  Embedding  &  0.761$\pm$0.003 (-0.111) & 0.790$\pm$0.003 & 0.793$\pm$0.006 (-0.045) & 0.835$\pm$0.006 & 0.827$\pm$0.003 (-0.031) & 0.904$\pm$0.003 \\
\midrule
              &  Projection  &  0.823$\pm$0.035 (-0.101) & 0.846$\pm$0.035 & 0.743$\pm$0.046 (-0.162) & 0.733$\pm$0.045 & 0.844$\pm$0.028 (-0.065) & 0.888$\pm$0.034 \\
              &  Prediction  &  0.875$\pm$0.004 (-0.049) & 0.903$\pm$0.001 & 0.869$\pm$0.002 (-0.036) & 0.898$\pm$0.003 & 0.871$\pm$0.002 (-0.038) & 0.924$\pm$0.003 \\
\multirow{-3}{*}{\textbf{Pubmed}}  &  Embedding  &  0.877$\pm$0.003 (-0.047) & 0.906$\pm$0.004 & 0.875$\pm$0.004 (-0.030) & 0.905$\pm$0.004 & 0.881$\pm$0.002 (-0.028) & 0.941$\pm$0.003 \\
\midrule
              &  Projection  &  0.685$\pm$0.022 (-0.225) & 0.668$\pm$0.024 & 0.691$\pm$0.014 (-0.220) & 0.667$\pm$0.012 & 0.700$\pm$0.013 (-0.219) & 0.707$\pm$0.010 \\
              &  Prediction  &  0.802$\pm$0.008 (-0.108) & 0.806$\pm$0.011 & 0.866$\pm$0.003 (-0.045) & 0.887$\pm$0.004 & 0.878$\pm$0.006 (-0.041) & 0.909$\pm$0.006 \\
\multirow{-3}{*}{\textbf{Citeseer}}  &  Embedding  &  0.804$\pm$0.008 (-0.106) & 0.811$\pm$0.008 & 0.877$\pm$0.004 (-0.034) & 0.898$\pm$0.005 & 0.883$\pm$0.006 (-0.036) & 0.914$\pm$0.006 \\
\midrule
              &  Projection  &  0.845$\pm$0.041 (-0.108) & 0.849$\pm$0.041 & 0.816$\pm$0.037 (-0.149) & 0.817$\pm$0.038 & 0.832$\pm$0.031 (-0.124) & 0.844$\pm$0.031 \\
              &  Prediction  &  0.942$\pm$0.002 (-0.011) & 0.948$\pm$0.002 & 0.950$\pm$0.002 (-0.015) & 0.955$\pm$0.002 & 0.950$\pm$0.001 (-0.006) & 0.974$\pm$0.001 \\
\multirow{-3}{*}{\textbf{Coauthor}}  &  Embedding  &  0.932$\pm$0.002 (-0.021) & 0.938$\pm$0.003 & 0.944$\pm$0.003 (-0.021) & 0.948$\pm$0.004 & 0.950$\pm$0.002 (-0.006) & 0.970$\pm$0.003 \\
\midrule
              &  Projection  &  0.867$\pm$0.021 (-0.062) & 0.868$\pm$0.032 & 0.887$\pm$0.016 (-0.048) & 0.903$\pm$0.009 & 0.905$\pm$0.006 (-0.032) & 0.936$\pm$0.011 \\
              &  Prediction  &  0.886$\pm$0.008 (-0.043) & 0.889$\pm$0.014 & 0.914$\pm$0.009 (-0.021) & 0.931$\pm$0.008 & 0.909$\pm$0.016 (-0.028) & 0.932$\pm$0.018 \\
\multirow{-3}{*}{\textbf{ACM}}  &  Embedding  &  0.884$\pm$0.004 (-0.045) & 0.874$\pm$0.011 & 0.874$\pm$0.009 (-0.061) & 0.888$\pm$0.011 & 0.909$\pm$0.009 (-0.028) & 0.941$\pm$0.006 \\
\midrule
              &  Projection  &  0.679$\pm$0.028 (-0.177) & 0.676$\pm$0.029 & 0.720$\pm$0.049 (-0.233) & 0.724$\pm$0.050 & 0.733$\pm$0.062 (-0.204) & 0.741$\pm$0.064 \\
              &  Prediction  &  0.811$\pm$0.029 (-0.045) & 0.778$\pm$0.027 & 0.925$\pm$0.008 (-0.028) & 0.931$\pm$0.006 & 0.899$\pm$0.021 (-0.038) & 0.916$\pm$0.023 \\
\multirow{-3}{*}{\textbf{Amazon}}  &  Embedding  &  0.890$\pm$0.005 (0.034) & 0.846$\pm$0.018 & 0.906$\pm$0.004 (-0.047) & 0.918$\pm$0.002 & 0.925$\pm$0.005 (-0.012) & 0.941$\pm$0.005 \\
\bottomrule
\end{tabular}
}
\label{tab:acc_fidelity_type1_sage}
\end{table*}

\begin{figure*}[ht]
\centering
\begin{subfigure}[t]{0.23\textwidth}
\includegraphics[width=\textwidth]{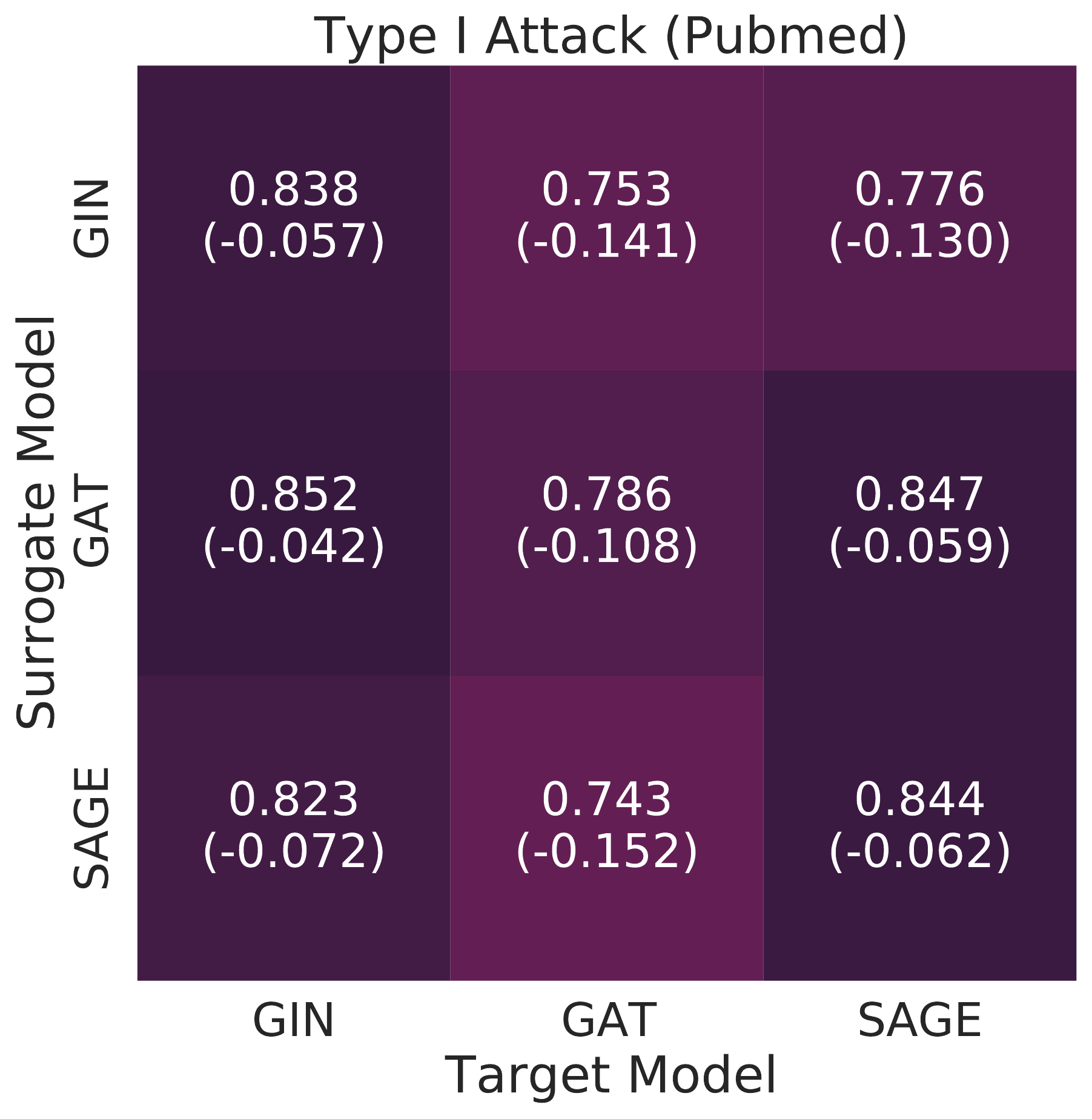}
\caption{t-SNE projection}
\label{fig:pubmed_projection_type1}
\end{subfigure}
\begin{subfigure}[t]{0.23\textwidth}
\includegraphics[width=\textwidth]{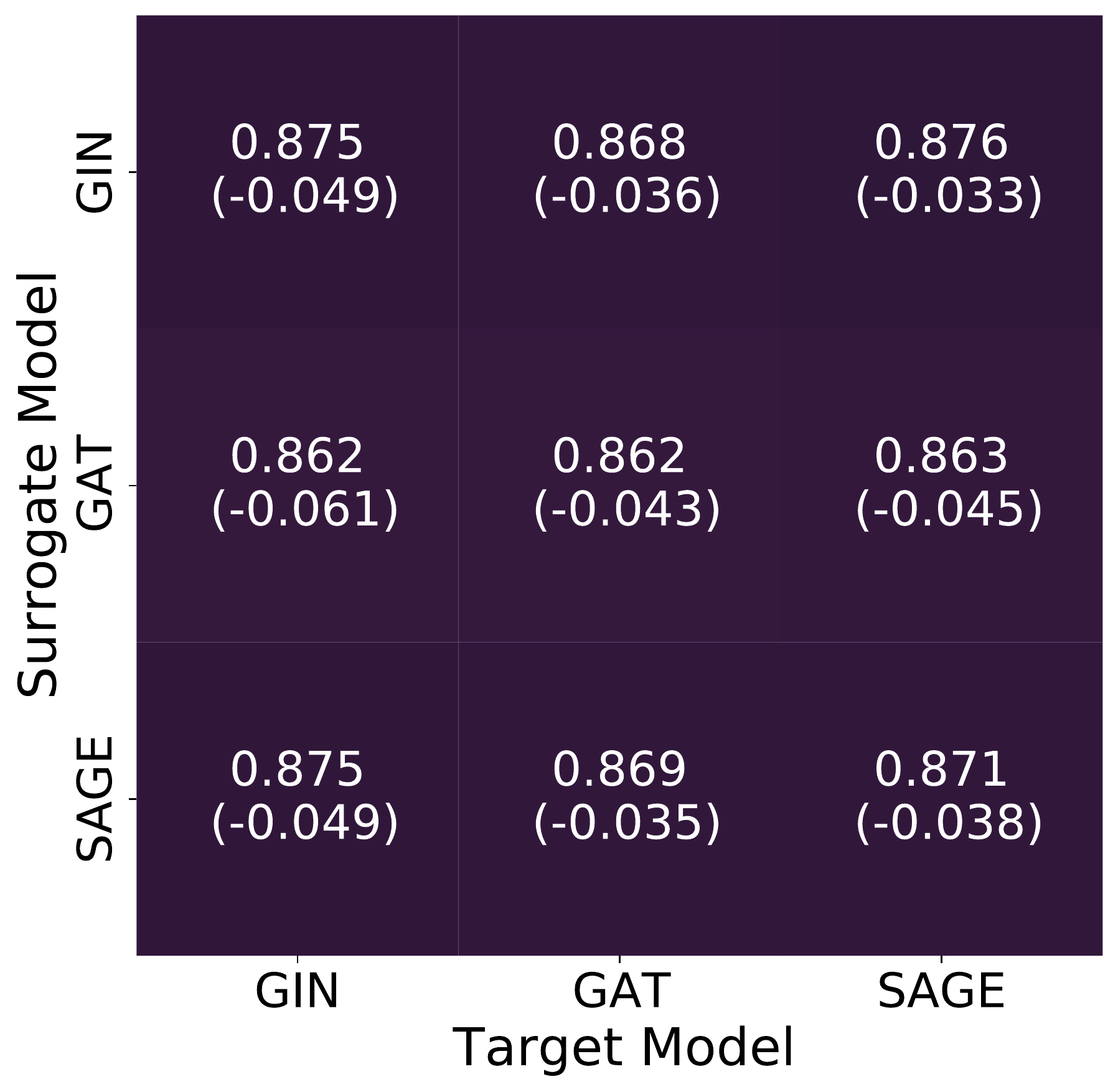}
\caption{Prediction}
\label{fig:pubmed_prediction_type1}
\end{subfigure}
\begin{subfigure}[t]{0.23\textwidth}
\includegraphics[width=\textwidth]{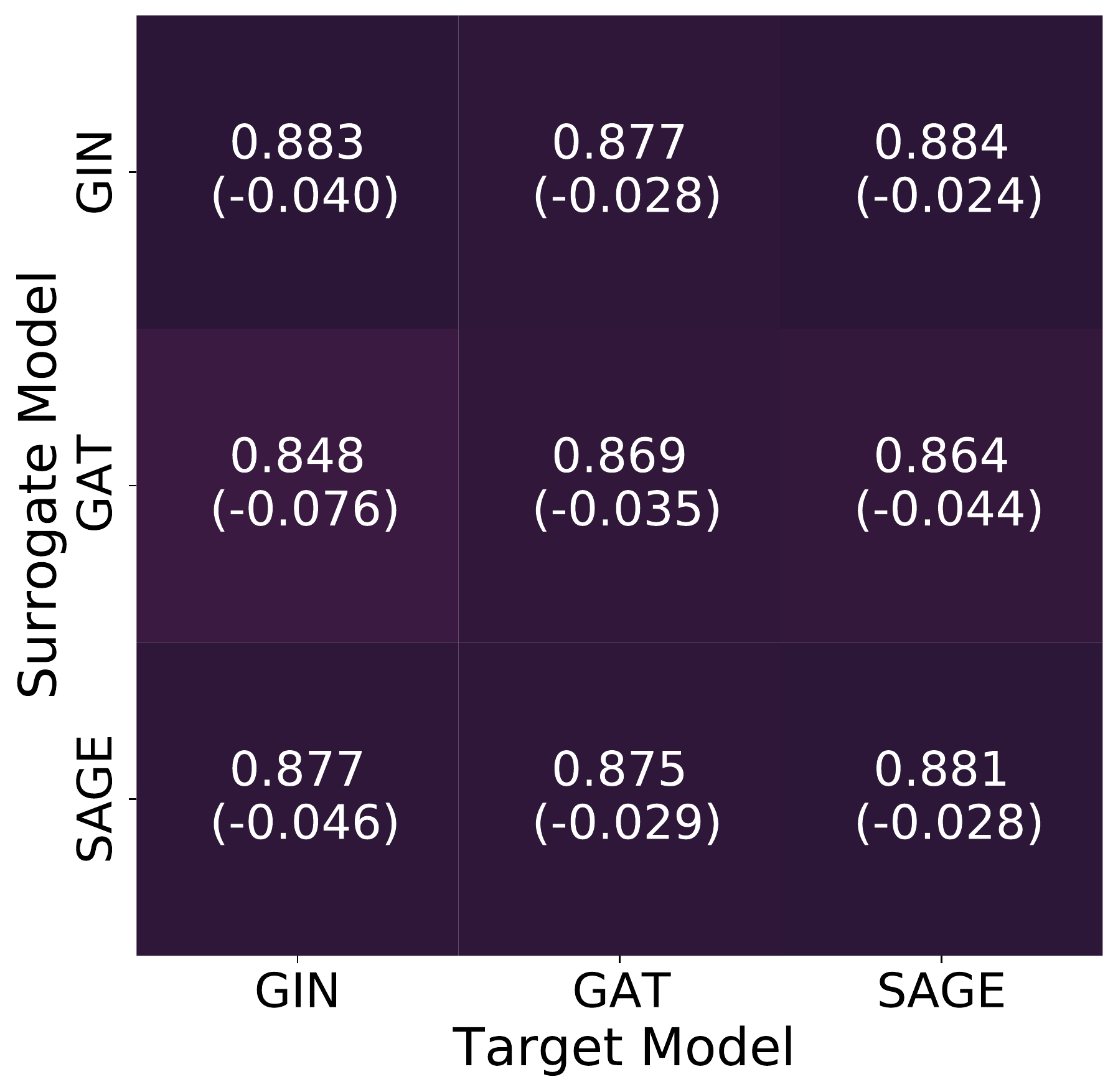}
\caption{Embedding}
\label{fig:pubmed_embedding_type1}
\end{subfigure}
\caption{Heatmap of the accuracy scores of Type I attacks. 
We show the performance results of 9 combinations of surrogate and target models given different response information. 
The accuracy differences (in parenthesis) of the surrogate models to the target models are also reported. 
We fix the dataset to Pubmed.}
\label{fig:pubmed_type1}
\end{figure*}

We first summarize the accuracy of the target models for the original node classification tasks in \autoref{tab:acc_target_models}.
We can observe that all GNN models achieve good performance on all datasets, which demonstrates that jointly considering node features and graph structure are effective for classification.
We then show the accuracy and fidelity of Type I attacks in \autoref{tab:acc_fidelity_type1_sage}.
Due to space limitations, we only show the attack results when the adversary uses GraphSAGE as the surrogate model. 
The performance results using GIN and GAT as the surrogate models follow similar patterns and can be found in \autoref{sec:type1_performance_appendix}.

\mypara{Accuracy}
As we can see from \autoref{tab:acc_fidelity_type1_sage}, Type I attacks can build surrogate models close to the target models given the response is predicted posterior probability or node embeddings (i.e., Type I.1 and Type I.2 attacks respectively). 
Take the Pubmed dataset as an example, the target models (i.e., GIN, GAT, and GraphSAGE) respectively achieve 0.924, 0.905, and 0.909 accuracy scores (see \autoref{tab:acc_target_models}), while the surrogate models can consistently achieve at least 0.869 accuracy score (see \autoref{tab:acc_fidelity_type1_sage}). 
This represents an approximately 0.04 accuracy score drop in all cases compared to the target models.
We can also observe that Type I attacks can build surrogate models that offer usable accuracy even the response is a 2-dimensional t-SNE projection matrix (i.e., Type I.3 attack). 
For instance, on the Pubmed dataset, the surrogate models achieve 0.823, 0.743, and 0.844 accuracy scores respectively.
This represents a 0.162 accuracy score drop in the worst case when the target model is GAT. 
For the rest of the five datasets, we also observe a subtle performance drop in Type I.3 attacks. 
Also, such performance drop, compared to Type I.1 and Type I.2 attacks, is expected since each t-SNE projection is only a 2-dimensional vector, which leads to additional information loss.
Overall, our results show that all three Type I attacks can build usable surrogate models.

Given our experimental configuration (i.e., three target models and three surrogate models), there are 9 different combinations for each response type given a single dataset.
Such configuration enables us to understand if the adversary can reliably steal target models in different circumstances.
Using the Pubmed dataset as an example, we summarize the attack accuracy performance in \autoref{fig:pubmed_type1}.
We can see that in general, the adversary can build accurate surrogate models given different combinations of GNN architectures for each response.
The subtle performance drop only occurs when GAT is the target model and the response is a t-SNE projection, i.e., Type I.3 attack.
Even in this case, the adversary can steal usable surrogate models with no more than a 0.162 accuracy score drop, which indicates our attack remains effective.
Our results demonstrate that the adversary does not require knowledge about the architecture of target models to conduct the attacks.

\begin{figure}[t]
\centering
\includegraphics[width=0.8\linewidth]{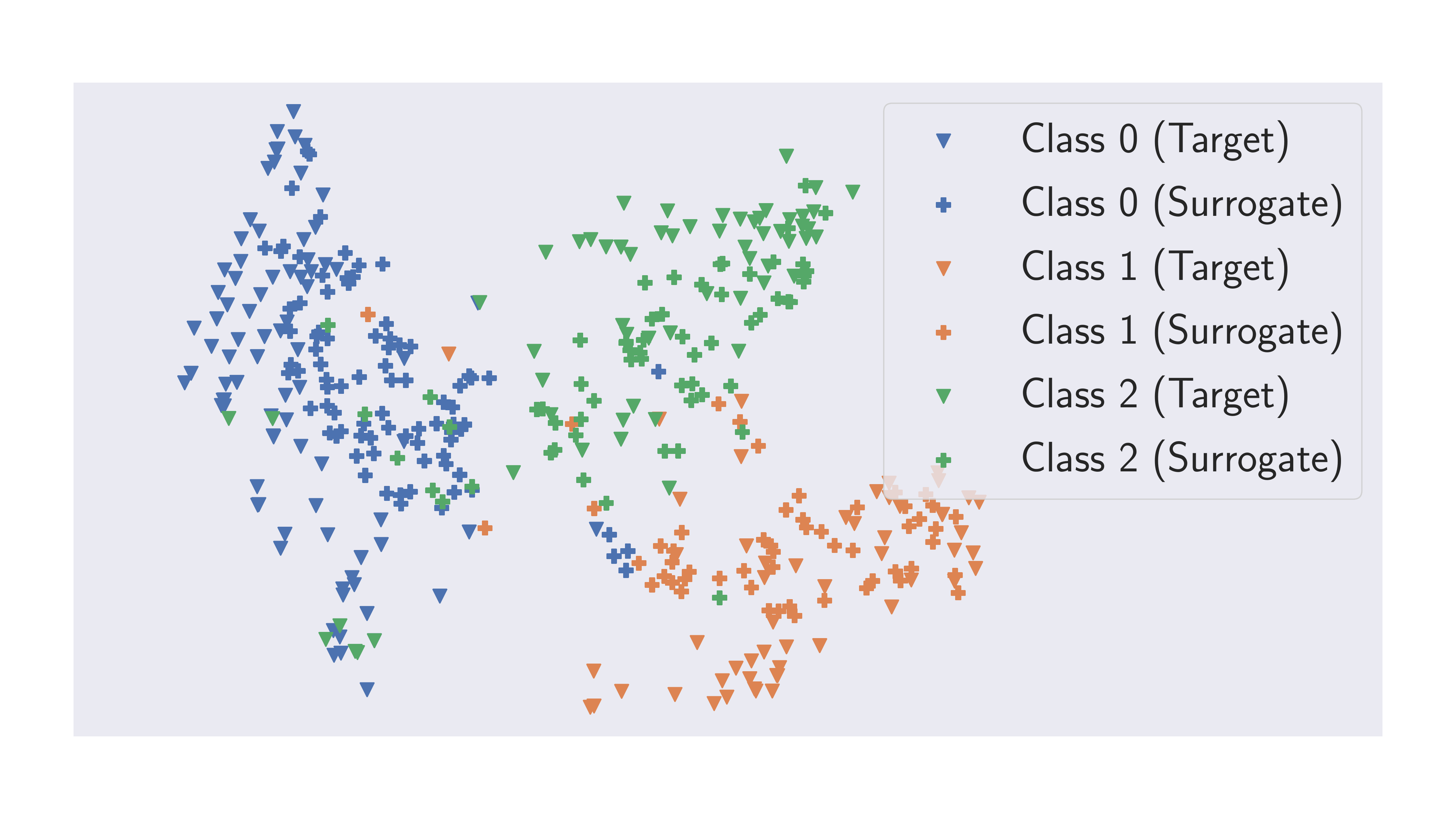}
\caption{The embeddings (256-dimension) obtained from the target and surrogate models of 200 randomly selected nodes on the ACM dataset. 
We project them into a 2-dimensional space using t-SNE. 
The target model is GAT and the surrogate model is SAGE.}
\label{fig:embedding_tsne_visualization}
\end{figure}

To better illustrate it, take the ACM dataset as an example, we extract the embeddings of a given set of nodes from both target and surrogate models and project them into a 2-dimensional space using t-SNE. 
The result is shown in \autoref{fig:embedding_tsne_visualization}.
We use the triangle (cross) to denote the embeddings extracted from the target (surrogate) model and different colors to denote different classes.
We find that for both target and surrogate models, the different classes' embeddings can be separated easily.
It means that the surrogate model can also successfully map nodes from different classes into different space, which lead to high accuracy.

\mypara{Fidelity}
The reconnaissance adversary's motivation is faithfully copying the behavior of the target model.
We also summarize the fidelity performance of the surrogate models in \autoref{tab:acc_fidelity_type1_sage}.
It is straightforward to see that the better accuracy performance a surrogate model can reach, the better fidelity it can achieve. 
For instance, on the Coauthor dataset, the surrogate models reach at least 0.816 accuracy score while these models achieve at least 0.817 fidelity score to the target models in all 9 cases.
The surrogate models for other datasets also follow similar patterns.
We then calculate the Pearson correlation coefficient between accuracy and fidelity of the surrogate models given three target models.
The coefficient scores are 0.957, 0.988, and 0.959 respectively.
The results exemplify that the fidelity of the surrogate models to the target models is highly correlated with their accuracy performance. 
From \autoref{fig:embedding_tsne_visualization}, we can also observe that for each class, the embeddings extracted from target and surrogate models lie in the same region, which implies that the surrogate model has the ability to generate the node embeddings that is close to the one generated by the target model, which leads to high fidelity.
Besides, we compare our attacks with Attack-3 proposed by Wu~et al.~\cite{WYPY20} using the overlapping Pubmed dataset and the results are shown in \autoref{tab:attack_comparison}.
Note that we use fewer data to train the surrogate model compared to Wu~et al.~\cite{WYPY20}. 
Given SAGE as $\mathcal{M}_S$ and GIN, GAT, and SAGE as $\mathcal{M}_T$, our attack achieves 0.903, 0.898, and 0.924 fidelity scores respectively while their attack only achieves 0.818 fidelity score.
The results exemplify that interacting with the target model can better facilitate the adversary to attain the reconnaissance goal.

\begin{table}[!t]
\centering
\caption{The accuracy and fidelity scores of Attack-3~\cite{WYPY20} and our Type I attacks on the Pubmed dataset. 
We use GraphSAGE as the surrogate model.}
\scalebox{0.9}{
\begin{tabular}{lcc}
\toprule
\textbf{Method} & \textbf{Accuracy} & \textbf{Fidelity}\\
\midrule
\textbf{Attack-3}~\cite{WYPY20} & 0.799 & 0.818 \\
\textbf{Our attack (GIN)} & 0.875 & 0.903 \\
\textbf{Our attack (GAT)} & 0.869 & 0.898 \\
\textbf{Our attack (SAGE)} & 0.871 & 0.924 \\
\bottomrule
\end{tabular}
}
\label{tab:attack_comparison}
\end{table}

\mypara{Stability}
We run each combination 5 times with different graph partition seeds. 
It enables us to measure how widely accuracy/fidelity values are dispersed from the average value (i.e., standard deviation).
A low standard deviation indicates a low volatility.
As we can observe in \autoref{tab:acc_fidelity_type1_sage}, the standard deviation values are low in all cases.
It shows that the adversary can steal from the target models with statistically stable accuracy and fidelity.

\mypara{Observation}
To achieve high fidelity, the adversary wants to make sure the mistakes and correct labels are the same  between the surrogate and target models. 
When the target model achieves high accuracy, it can make the correct predictions for most of the test data while making few mistakes. 
If the surrogate model gets close to the accuracy performance of the target model, it would gain high fidelity to the target model due to the fact the number of correctly predicted instances by the surrogate model would considerably overlap those made by the target model. 
The above Pearson correlation results verify this intuitive explanation.

\mypara{Takeaways} 
The theft adversary can reliably build accurate surrogate models close to the target models via Type I attacks.
It is worth noting that, in the real world, it is hard for the reconnaissance adversary to comprehensively verify the fidelity of the surrogate models without risking a number of queries. 
Our results imply that the reconnaissance adversary may focus on building surrogate models that preserve high accuracy similar to the target models.
In turn, these surrogate models would likely be faithful to the remote targets. 
Besides, our results demonstrate that the adversary does not require knowledge of the target models' architectures.

% ----------------------------------------------------
\subsection{Performance Evaluation: Type II Attacks}
\label{sec:type2_performance}
% ----------------------------------------------------

\begin{figure}[t]
\centering
\includegraphics[width=0.8\linewidth]{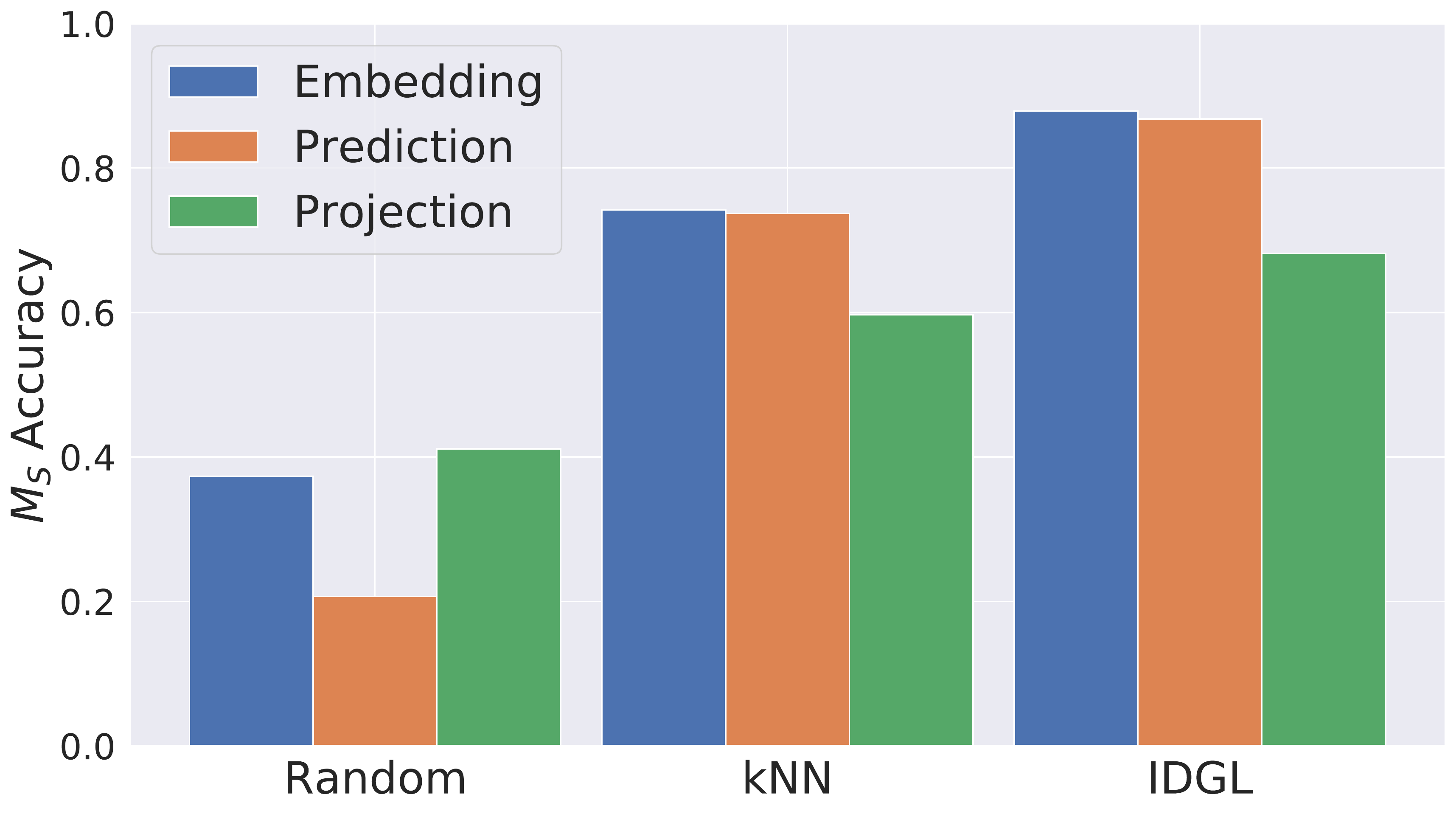}
\caption{The accuracy scores of Type II attacks on the Citeseer dataset using different graph reconstruction methods. 
We use GAT as the target model and GraphSAGE as the surrogate model.
The fidelity scores can be found in \autoref{sec:random_knn_idgl_comparison_fidelity}.}
\label{fig:graph_reconstruction_comparison}
\end{figure}

\begin{table*}[!t]
\centering
\caption{The accuracy and fidelity scores of Type II attacks using different response information on all the 6 datasets. 
Both average values and standard deviations are reported. 
The accuracy differences (in parenthesis) of the surrogate models to the target models are also reported. 
We use GraphSAGE as the surrogate model.}
\scalebox{0.8}{
\begin{tabular}{lccccccc}
\toprule
\multirow{3}{*}{\textbf{Dataset}} &
  \multirow{3}{*}{\begin{tabular}[x]{@{}c@{}}$\mathcal{M}_S$\\(SAGE)\end{tabular}} &
  \multicolumn{6}{c}{$\mathcal{M}_T$} \\
\cmidrule{3-8}
 &
   &
  \multicolumn{2}{c}{\textbf{GIN}} &
  \multicolumn{2}{c}{\textbf{GAT}} &
  \multicolumn{2}{c}{\textbf{SAGE}} \\
 \cmidrule(lr){3-4} \cmidrule(lr){5-6} \cmidrule(lr){7-8}
 &
   &
  \textbf{Accuracy} &
  \textbf{Fidelity} &
  \textbf{Accuracy} &
  \textbf{Fidelity} &
  \textbf{Accuracy} &
  \textbf{Fidelity} \\

\midrule
              &  Projection  &  0.703$\pm$0.018 (-0.169) & 0.732$\pm$0.018 & 0.675$\pm$0.007 (-0.163) & 0.693$\pm$0.009 & 0.713$\pm$0.024 (-0.145) & 0.761$\pm$0.028 \\
              &  Prediction  &  0.779$\pm$0.008 (-0.093) & 0.824$\pm$0.009 & 0.782$\pm$0.005 (-0.056) & 0.832$\pm$0.004 & 0.809$\pm$0.004 (-0.049) & 0.882$\pm$0.004 \\
\multirow{-3}{*}{\textbf{DBLP}}  &  Embedding  &  0.783$\pm$0.009 (-0.089) & 0.825$\pm$0.010 & 0.787$\pm$0.004 (-0.051) & 0.834$\pm$0.009 & 0.812$\pm$0.004 (-0.046) & 0.886$\pm$0.003 \\
\midrule
              &  Projection  &  0.836$\pm$0.015 (-0.088) & 0.862$\pm$0.015 & 0.739$\pm$0.018 (-0.166) & 0.738$\pm$0.022 & 0.850$\pm$0.017 (-0.059) & 0.893$\pm$0.020 \\
              &  Prediction  &  0.878$\pm$0.006 (-0.046) & 0.908$\pm$0.003 & 0.867$\pm$0.003 (-0.038) & 0.897$\pm$0.003 & 0.872$\pm$0.004 (-0.037) & 0.928$\pm$0.004 \\
\multirow{-3}{*}{\textbf{Pubmed}}  &  Embedding  &  0.879$\pm$0.004 (-0.045) & 0.910$\pm$0.005 & 0.875$\pm$0.003 (-0.030) & 0.905$\pm$0.002 & 0.881$\pm$0.002 (-0.028) & 0.941$\pm$0.002 \\
\midrule
              &  Projection  &  0.647$\pm$0.028 (-0.263) & 0.658$\pm$0.024 & 0.682$\pm$0.030 (-0.229) & 0.679$\pm$0.033 & 0.695$\pm$0.021 (-0.224) & 0.712$\pm$0.022 \\
              &  Prediction  &  0.834$\pm$0.012 (-0.076) & 0.862$\pm$0.011 & 0.868$\pm$0.005 (-0.043) & 0.896$\pm$0.004 & 0.880$\pm$0.007 (-0.039) & 0.928$\pm$0.007 \\
\multirow{-3}{*}{\textbf{Citeseer}}  &  Embedding  &  0.827$\pm$0.009 (-0.083) & 0.850$\pm$0.014 & 0.879$\pm$0.006 (-0.032) & 0.904$\pm$0.008 & 0.880$\pm$0.006 (-0.039) & 0.923$\pm$0.010 \\
\midrule
              &  Projection  &  0.862$\pm$0.052 (-0.091) & 0.866$\pm$0.054 & 0.864$\pm$0.044 (-0.101) & 0.867$\pm$0.043 & 0.793$\pm$0.044 (-0.163) & 0.801$\pm$0.046 \\
              &  Prediction  &  0.955$\pm$0.001 (0.002) & 0.960$\pm$0.001 & 0.954$\pm$0.001 (-0.011) & 0.958$\pm$0.001 & 0.954$\pm$0.002 (-0.002) & 0.976$\pm$0.002 \\
\multirow{-3}{*}{\textbf{Coauthor}}  &  Embedding  &  0.950$\pm$0.002 (-0.003) & 0.955$\pm$0.001 & 0.947$\pm$0.003 (-0.018) & 0.950$\pm$0.002 & 0.951$\pm$0.002 (-0.005) & 0.965$\pm$0.006 \\
\midrule
              &  Projection  &  0.854$\pm$0.036 (-0.075) & 0.842$\pm$0.035 & 0.888$\pm$0.012 (-0.047) & 0.904$\pm$0.012 & 0.918$\pm$0.007 (-0.019) & 0.947$\pm$0.008 \\
              &  Prediction  &  0.911$\pm$0.007 (-0.018) & 0.899$\pm$0.018 & 0.927$\pm$0.004 (-0.008) & 0.934$\pm$0.006 & 0.923$\pm$0.005 (-0.014) & 0.944$\pm$0.007 \\
\multirow{-3}{*}{\textbf{ACM}}  &  Embedding  &  0.887$\pm$0.012 (-0.042) & 0.877$\pm$0.010 & 0.876$\pm$0.005 (-0.059) & 0.889$\pm$0.012 & 0.921$\pm$0.009 (-0.016) & 0.944$\pm$0.010 \\
\midrule
              &  Projection  &  0.640$\pm$0.020 (-0.216) & 0.640$\pm$0.021 & 0.691$\pm$0.015 (-0.262) & 0.662$\pm$0.019 & 0.739$\pm$0.034 (-0.198) & 0.743$\pm$0.033 \\
              &  Prediction  &  0.802$\pm$0.031 (-0.054) & 0.767$\pm$0.026 & 0.884$\pm$0.005 (-0.069) & 0.890$\pm$0.008 & 0.922$\pm$0.009 (-0.015) & 0.931$\pm$0.007 \\
\multirow{-3}{*}{\textbf{Amazon}}  &  Embedding  &  0.741$\pm$0.109 (-0.115) & 0.734$\pm$0.117 & 0.892$\pm$0.018 (-0.061) & 0.900$\pm$0.019 & 0.938$\pm$0.005 (0.001) & 0.954$\pm$0.005 \\
\bottomrule
\end{tabular}
}
\label{tab:acc_fidelity_type2_sage}
\end{table*}

\begin{figure*}
\centering
\begin{subfigure}[t]{0.23\textwidth}
\includegraphics[width=\textwidth]{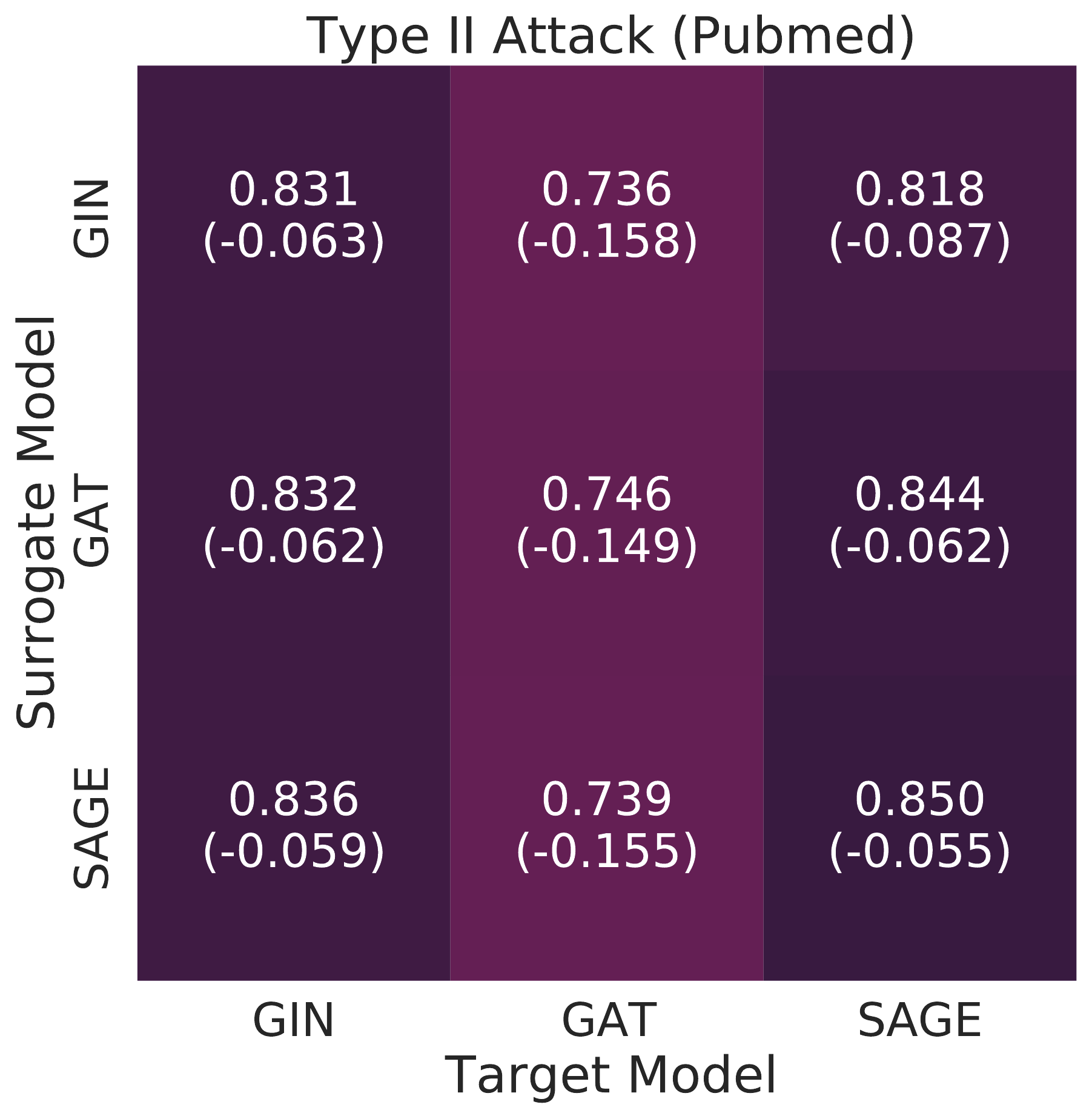}
\caption{t-SNE Projection} 
\label{fig:pubmed_projection_type2}
\end{subfigure}
\begin{subfigure}[t]{0.23\textwidth}
\includegraphics[width=\textwidth]{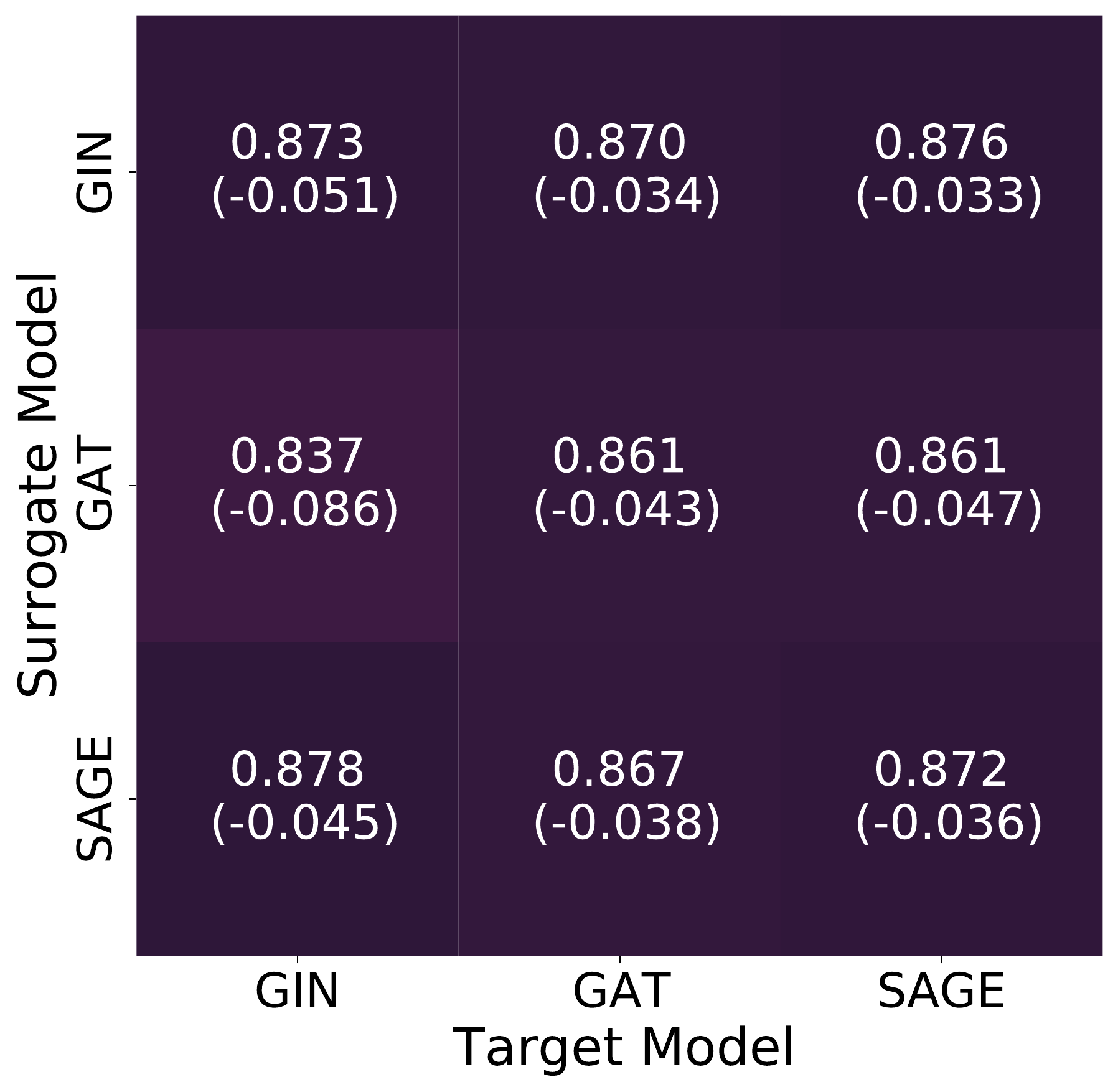}
\caption{Prediction} 
\label{fig:pubmed_prediction_type2}
\end{subfigure}
\begin{subfigure}[t]{0.23\textwidth}
\includegraphics[width=\textwidth]{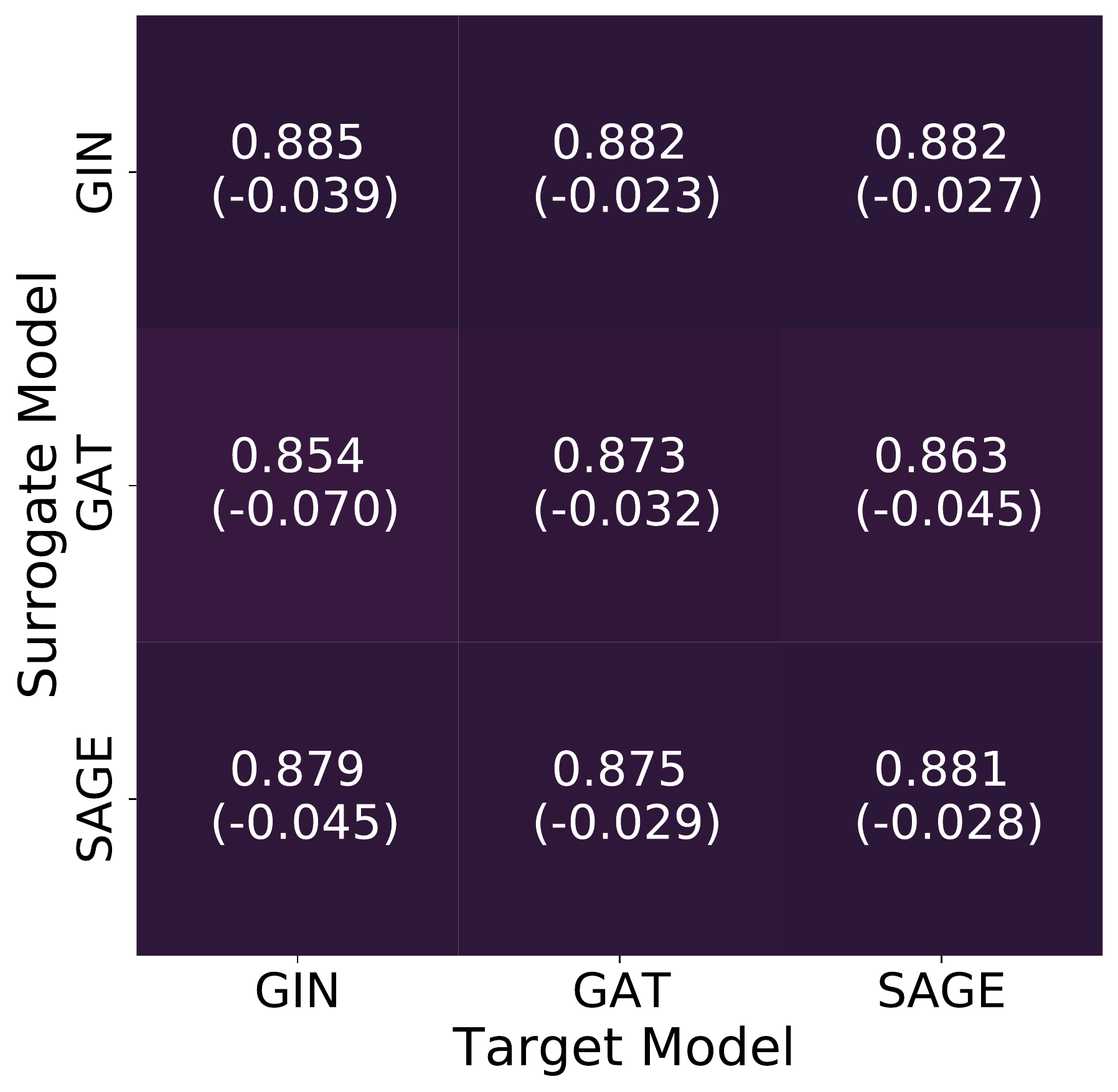}
\caption{Embedding} 
\label{fig:pubmed_embedding_type2}
\end{subfigure}
\caption{Heatmap of the accuracy scores of Type II attacks. 
We show the performance results of 9 combinations of surrogate and target models given different response information. 
The accuracy differences (in parenthesis) of the surrogate models to the target models are also reported. 
We fix the dataset to Pubmed.}
\label{fig:pubmed_type2}
\end{figure*}

To launch Type II attacks, the adversary first builds an adjacency matrix $\mathbf{A}_Q$ for query graph $\mathbf{G}_Q$ before querying the target model $\mathcal{M}_T$.
We follow the aforementioned graph reconstruction configuration to restore $\mathbf{A}_Q$ for query graph $\mathbf{G}_Q$ and conduct the model stealing attacks.
The performance of Type II attacks is summarized in \autoref{tab:acc_fidelity_type2_sage}.
Due to space limitations, we only show the attack results when the adversary uses GraphSAGE as the surrogate model. 
The performance results using GIN and GAT as the surrogate models follow similar patterns and can be found in \autoref{sec:type2_performance_appendix}.

\mypara{Accuracy}
As we can see in \autoref{tab:acc_fidelity_type2_sage}, given all the datasets, the adversary can launch Type II.1/2 attacks to build surrogate models that offer accuracy on par with the target models. 
At the same time, we observe that the adversary can launch Type II.3 attacks and steal usable surrogate models.  
Take the Pubmed dataset as an example, when the response is the t-SNE projection of query nodes, the surrogate models achieve the average accuracy score of 0.836, 0.739, and 0.850 concerning different target models.
This represents  0.166 accuracy drop in the worst case (compared to the target model performance). 
In general, Type II.3 attacks can achieve comparable accuracy performance in all 6 datasets.
Also, we investigate if the adversary can build accurate surrogate models given different combinations of GNN architectures for each response in Type II attacks.
As we can observe from \autoref{fig:pubmed_type2}, when the response is predicted posterior probability or embedding, the accuracy of the surrogate model is close to the target model.
Regarding the case when the response is t-SNE projection,  our  attack  still  works  well. 
We only witness a slight performance drop when the target model is GAT. 
The results exemplify that, in general, Type II attacks remain effective in stealing target models with different architectures.

\mypara{Fidelity}
We also observe the same correlation between accuracy and fidelity in Type II attacks. 
That is, the better accuracy performance a surrogate model can reach, the better fidelity it can achieve. 
Take the Coauthor dataset as an example, the surrogate models reach at least 0.793 accuracy score while these models achieve at least 0.801 fidelity score to the target models in all cases.
We also calculate the Pearson correlation coefficient between accuracy and fidelity of the surrogate models given three target models.
The coefficient scores are 0.970, 0.991, and 0.967 respectively.
These correlation scores are similar to what we observe from Type I attack results. 

\mypara{Efficacy of Learned Query Graph Structure}
To investigate the effect of graph structure, we compare IDGL with two additional methods, i.e., random graph construction and $k$NN.
For the two additional methods, we set the average node degree to 24, which is the same value used to initialize IDGL.
Due to space limitations, we only show the accuracy of Type II attacks on the Citeseer dataset using GAT as the target model and GraphSAGE as the surrogate model.
The accuracy and fidelity of other datasets follow similar patterns.
As we can see from \autoref{fig:graph_reconstruction_comparison}, using IDGL to reconstruct the graph structure reaches the highest attack accuracy for all responses.
For instance, the accuracy score is 0.879 when using IDGL as the reconstruction method and taking embedding as the response, while the corresponding accuracy score is only 0.411 and 0.737 respectively when using random graph construction or $k$NN as the reconstruction method.
It demonstrates that an effective graph reconstruction method does benefit the final attack performance.

\mypara{Stability}
As we can observe in \autoref{tab:acc_fidelity_type2_sage}, the standard deviation values remain low in all cases.
This shows that the adversary can steal the target models with statistically stable accuracy and fidelity in Type II attacks.

\mypara{Observation}
When comparing \autoref{tab:acc_fidelity_type2_sage} to \autoref{tab:acc_fidelity_type1_sage}, we observe that Type II attack achieves better performance than Type I attack in certain cases. 
Chen~et al.~\cite{CWZ20} observed a similar phenomenon and concluded that the raw graphs are not always optimal for the downstream tasks for different reasons. 
For example, raw graphs may contain noisy/incomplete information due to the error-prone data collection or their structures do not reflect the ideal graph topology after feature extraction and transformation.
The query graphs learned by the IDGL framework in our Type II attacks are optimized toward the downstream tasks (e.g., node classification) and may achieve better performance in some cases.  
We refer the audience to Chen~et al.~\cite{CWZ20} for additional details.

\mypara{Takeaways}
Our results show that the attack framework enables the adversary to learn a discrete graph structure and steal usable surrogate models. 
Coupling with the results shown in \autoref{sec:type1_performance}, we demonstrate that our model stealing attacks achieve strong performance concerning different responses.

% ----------------------------------------------------
\subsection{Performance Evaluation: Query Budget}
\label{sec:budget}
% ----------------------------------------------------

\begin{figure}[t]
\centering
\includegraphics[width=0.85\linewidth]{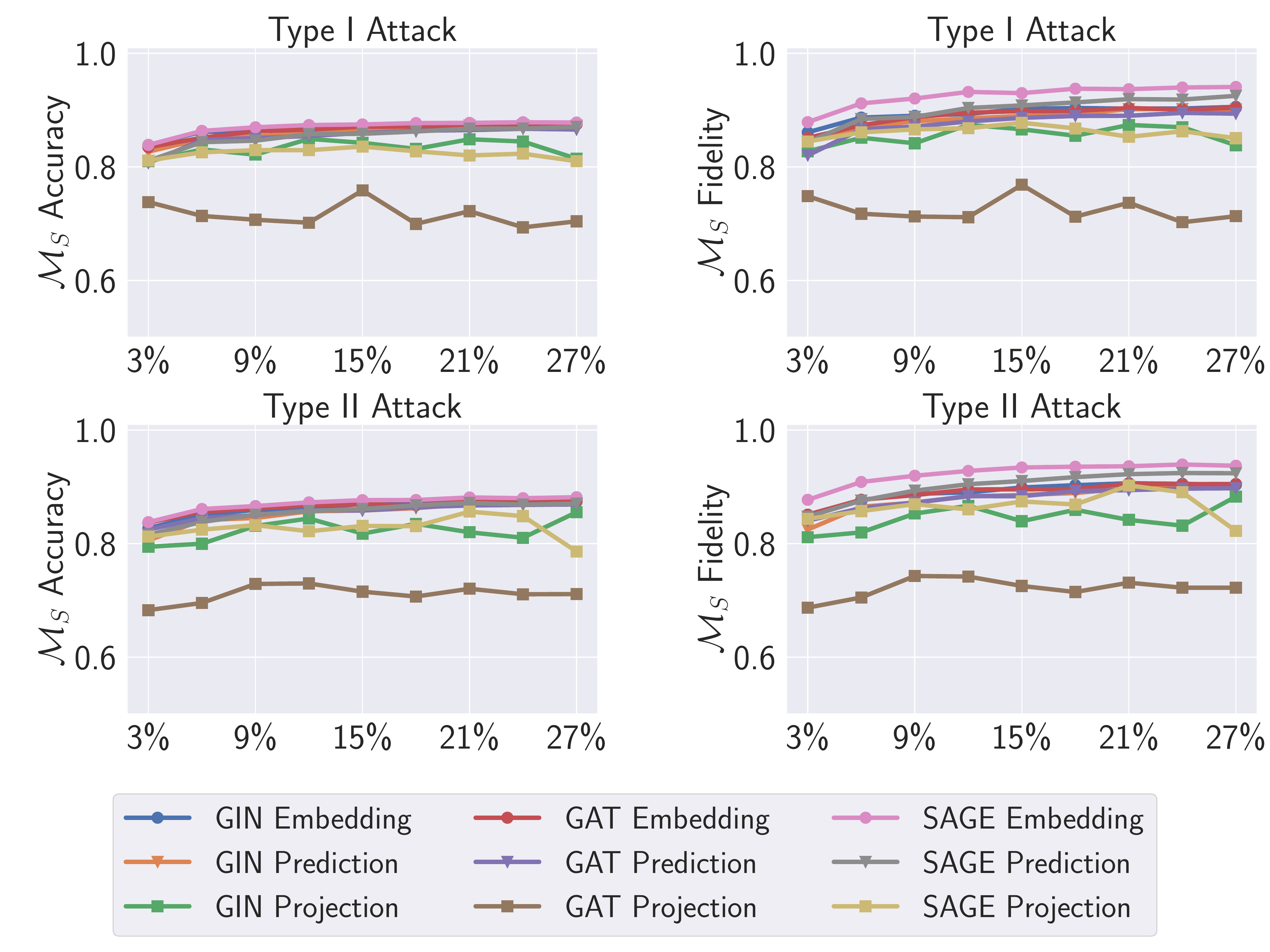}
\caption{The average accuracy (fidelity) scores of Type I (Type II) attacks using different response information on the Pubmed datasets. 
We use GraphSAGE as the surrogate model. The x-axis represents the percentage of randomly selected nodes in the dataset and the y-axis represents the accuracy (fidelity) of surrogate models.}
\label{fig:pubmed_query_budget}
\end{figure}

We then investigate the attack performance with respect to different query budgets, i.e., different sizes of query graph $\mathbf{G}_Q$.
Due to space limitations, we only show the results for the Pubmed dataset, other datasets follow similar trends.
The corresponding accuracy and fidelity of Type I and II attacks are summarized in \autoref{fig:pubmed_query_budget}.
We observe that in general, larger query budgets lead to better accuracy and fidelity.
For instance, in Type I attack, when the response is GraphSAGE's embedding, the accuracy score increases from 0.839 to 0.870 when the query budget increases from 3\% to 27\%.
However, in most of the cases, we can achieve similar performance even using only 3\% of randomly sampled nodes of the original dataset (10\% of the $\mathbf{G}_Q$ used in previous experiments).
The results demonstrate that the adversary can still launch effective attacks even with a low-quality query graph (less query budget and no graph structure information).
Besides, when the query budget is small, the adversary can investigate the characteristics of query data (e.g., sparsity of the learned graph) and decide if edge reconstruction is necessary to launch attacks.

% ----------------------------------------------------
\section{Discussions}
\label{sec:discussion}
% ----------------------------------------------------

\begin{figure}[t]
\centering
\begin{subfigure}[t]{0.49\linewidth}
\includegraphics[width=\textwidth]{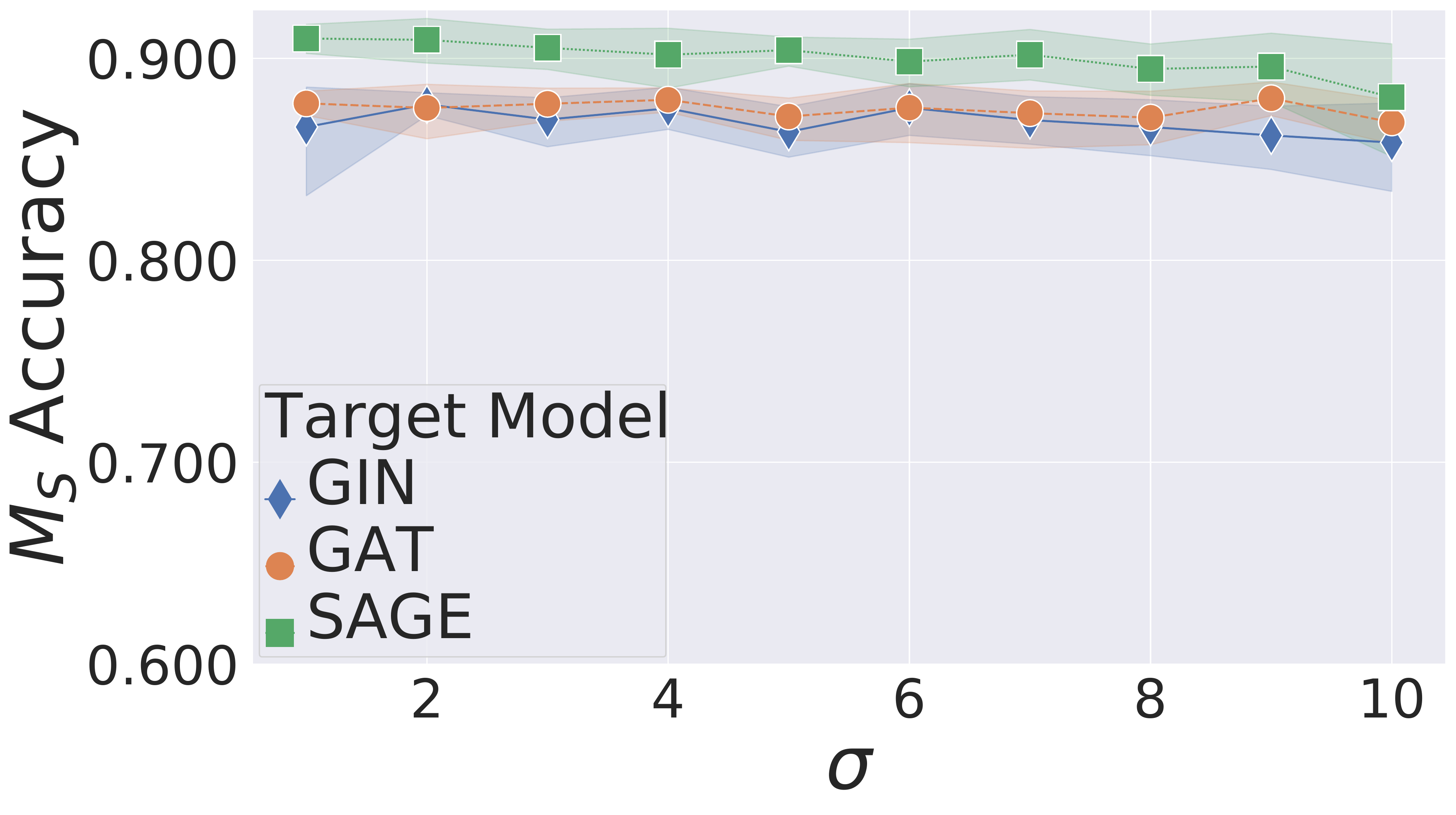}
\caption{t-SNE Projection} 
\label{fig:acm_tsne_noise}
\end{subfigure}
\centering
\begin{subfigure}[t]{0.49\linewidth}
\includegraphics[width=\textwidth]{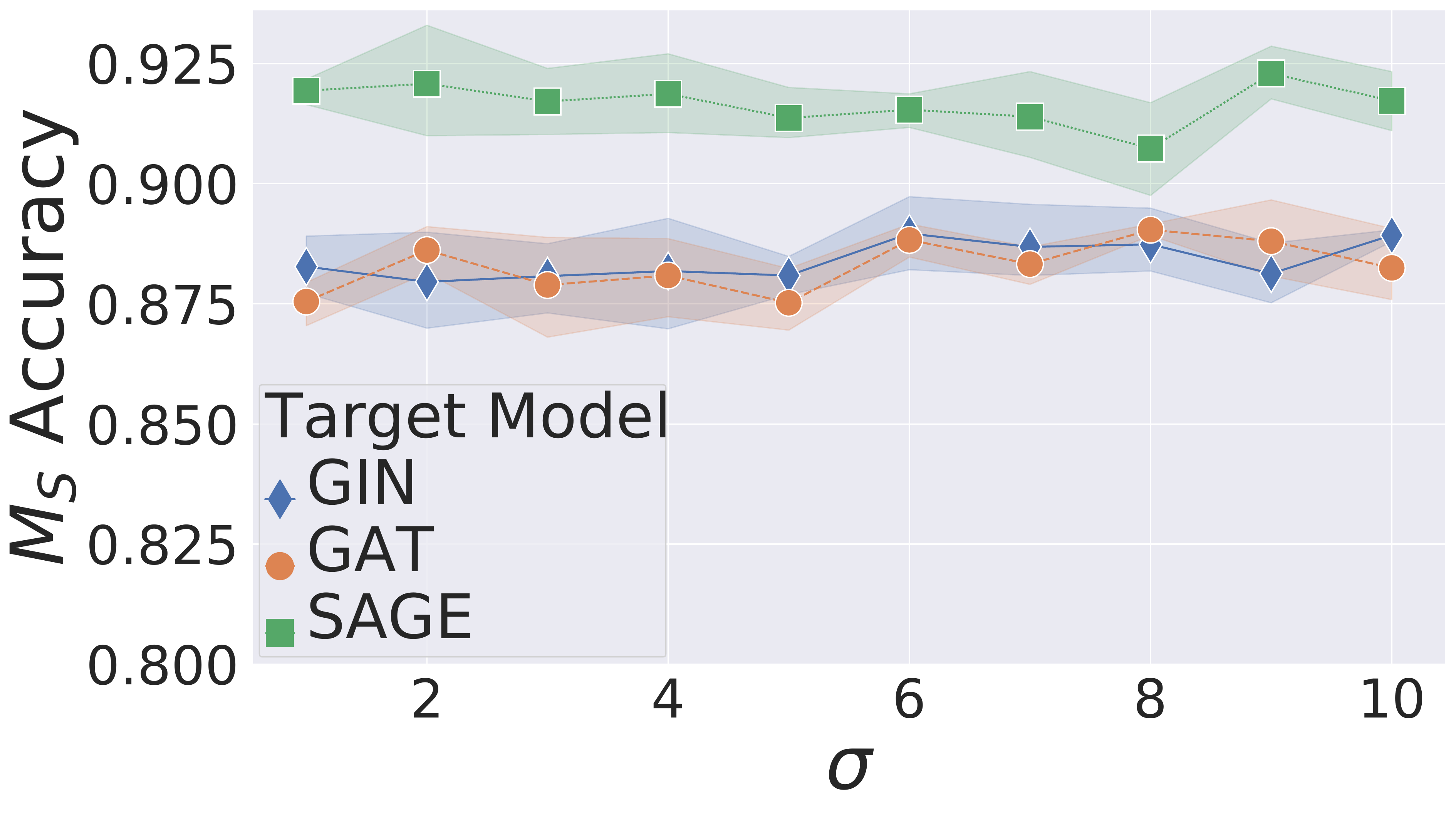}
\caption{Embedding} 
\label{fig:acm_embedding_noise}
\end{subfigure}
\caption{Defense: adding Gaussian noise to t-SNE projection and node embedding. 
We report the accuracy scores and standard deviation of the surrogate model (GraphSAGE) given all three target models under this defense mechanism (Type I.1 and I.3 attacks, respectively). 
$\sigma$ is the standard deviation of the Gaussian noise. 
We fix the dataset to ACM.}
\label{fig:defense_noise}
\end{figure}

\mypara{Limitation}
Our model stealing attack is limited to the scope that the target model returns node-level results.
We did not explore the scenario when the target model accepts an arbitrary graph as input and returns graph-level results.
That is, the target models represent the whole structure of graphs using various pooling methods and return a single embedding vector for downstream graph-level tasks such as graph classification~\cite{XPJW21,XHLJ19}.
Model stealing attack under this setting would require a different approach.
Besides, we do not jointly optimize both graph structure and surrogate model. 
Such training paradigm may cost more training epochs to converge, and the query budget would also increase, which inevitably increases the risk of being detected.
We put them into our future work.

\mypara{Defense}
Several countermeasures to model stealing attacks have been discussed in previous literature~\cite{TZJRR16,OSF19}.
One straightforward countermeasure is injecting perturbations to the predicted posterior probability reported by the classifier, i.e., perturb the probability while retaining the top-1 label~\cite{TZJRR16,LEMS19}.
To cope with different types of query responses, we consider adding random noise to the response regardless of the corresponding label, i.e. the distribution of the random noise is independent of the node class label.
Concretely, we add random Gaussian noise into node embedding and t-SNE projection returned by all three target models and use GraphSAGE as the surrogate model to understand the effectiveness of such countermeasure. According to the defined threat model, the accuracy of the surrogate model measures the attack performances. 
A higher accuracy of the surrogate model indicates a more successful attack. 
We use the ACM dataset as an example and the results are shown in \autoref{fig:defense_noise}.
We observe that the random Gaussian noise slightly affects the accuracy performance of the surrogate model. 
For instance, in \autoref{fig:acm_tsne_noise}, when $\sigma$ (i.e., the standard deviation of the added noise) is greater than 7, we can observe the accuracy performance of the surrogate model starts to decrease for different target models. In contrast, if the query response from the target model is the node embedding vector, we can only observe much fewer fluctuations of the surrogate model's accuracy with increasingly stronger random Gaussian noise injected to the embedding (see \autoref{fig:acm_embedding_noise}). 
To summarize, our preliminary experiment does not establish concrete evidence that adding random noise would counter our attacks.
We leave designing effective defense mechanisms for model stealing attacks against GNNs as our future work. 

% ----------------------------------------------------
\section{Related Work}
\label{sec:related_work}
% ----------------------------------------------------

In this section, we review the research work close to our proposed attacks.
We refer the readers to~\cite{GF18,ZCZYLS18,LRKAK19,ZCZ20,WPCLZY20} for an in-depth overview of different GNN models, and~\cite{SDYWYHL18,DLTHWZS18,CLPXCXHZ20,JLXWT20,XMLDLTJ20} for comprehensive surveys of existing adversarial attacks and defense strategies on GNNs.

\mypara{Model Stealing Attack Against ML Models}
Model extraction is in many ways similar to model distillation, but it differs in that the victim’s proprietary training set is not accessible to the adversary. 
In this regard, previous literature already investigated stealing various aspects of a black-box ML model such as hyperparameters~\cite{WG18}, architecture~\cite{OASF18}, information on training data~\cite{HJBGZ21,HWWBSZ21}, parameters~\cite{TZJRR16,CJM20}, decision boundaries~\cite{PMGJCS17}, and functionality~\cite{JCBKP20,OSF19}.
However, most of these efforts focused on images.
There exist some preliminary work on model stealing attacks against GNNs~\cite{DR19,WYPY20}.
However, they are only focusing on the transductive setting of GNNs, which cannot generalize to unseen data.
Our attacks instead focus on a more popular and general setting of GNNs, i.e., inductive setting. 
We fill the gap and understand if both theft and reconnaissance adversaries can steal inductive GNNs with high accuracy and high fidelity.

\mypara{Causative Attacks on GNNs} 
Many adversarial attacks to GNNs~\cite{BG192,EADP20,XCLCWHL19,ZG19,SWTHH20,MDM20,STLLXCS18,DLTHWZS18,ZJWG20,ZCBSZ22} are causative attacks~\cite{HJNRT11}.
These attacks assume that an adversary can manipulate the training dataset in order to change the parameters of the target model and influence their behavior. 
In this context, Z{\"u}gner~et al.~\cite{ZAG18} was the first to introduce unnoticeable adversarial perturbations of the node’s features and the graph structure. 
Their goal was to reduce the accuracy of node classification via GCN. 
After this work, different adversarial attack strategies have been proposed.
Depending on the attack objectives, they aim at 
reducing the accuracy of node classification~\cite{BG192,CHLW20,XCLCWHL19,ZG19,SWTHH20,MDM20} (node level), link prediction~\cite{STLLXCS18,BG192,CLSL20,LJL20} (edge level), graph classification~\cite{DLTHWZS18,ZJWG20} (graph level), etc.
Our attack does not tamper with the training graph data and does not change the behavior of the target model or its parameters.

\mypara{Exploratory Attacks on GNNs} 
In reality, however, it is more practical for the attacker to query the target model and leverage the model’s responses on these carefully crafted input data.
Consequently, we witness the emerging of exploratory attacks on ML models. 
However, adversarial exploratory attacks on GNNs remain understudied.
In particular, only a few studies~\cite{HJBGZ21,WYPY20,DBS20} focused on exploratory attacks on GNNs.
For instance, He~et al.~\cite{HJBGZ21} proposed the first link stealing attack to infer if there exists an edge between a given pair of nodes in the training graph. 
Duddu~et al.~\cite{DBS20} and He~et al.~\cite{HWWBSZ21} discussed the membership inference attack that infers whether a given node in the graph was used to train the target model by leveraging different background knowledge.
Note that, in membership inference attacks, the attack model in its training phase does not interact with the target model. However, in model stealing attacks, the surrogate model does have interaction with the target model since it needs the guidance of the target model to optimize its parameters.

\mypara{Defense of Attacks on GNNs}
To mitigate those attacks, several defense strategies have been proposed.
The core idea of the existing defense strategies is reducing the sensitivity of GNNs using adversarial training~\cite{DSZLW19,FHTC19,DDZ19,SLGZ19,WLH19,JZ20}, perturbation detection~\cite{IBG19}, graph sanitization~\cite{ZKC19,WWTDLZ19}, etc.
In turn, the trained GNNs are robust to perturbation (e.g., structure perturbation~\cite{DSZLW19,WLH19,JZ20}, attribution perturbation~\cite{FHTC19,DDZ19,SLGZ19,WLH19,JZ20}).
Also, robustness certification~\cite{BG19,ZG192} becomes an emerging research direction.
They aim at reasoning the safety posture of GNNs under adversarial perturbations.
However, these defense techniques only protect GNNs from causative attacks instead of exploratory attacks.

% ----------------------------------------------------
\section{Conclusion}
\label{sec:conclusion}
% ----------------------------------------------------

In this paper, we perform the first security risk assessment against inductive GNNs through the lens of model stealing attacks.
We propose a threat model to systematically categorize an adversary's background knowledge into two dimensions, i.e., query graph and model responses.
By jointly considering the two dimensions, we summarize six attack scenarios.
We then propose a general attack framework that can be applied in different scenarios.
Extensive experiments on three popular inductive GNN architectures and six benchmark datasets show that our model stealing attacks can handle different types of responses and achieve strong performance.
Moreover, the attacks are still effective even the adversary has no knowledge about the graph structural information.

\section*{Acknowledgement}
We thank the anonymous shepherd and reviewers for their feedback in improving this paper.
This work is partially funded by the Helmholtz Association within the project ``Trustworthy Federated Data Analytics'' (TFDA) (funding number ZT-I-OO1 4).
This work is also supported by the Helmholtz Association's Initiative and Networking Fund on the HAICORE@FZJ partition.

% ----------------------------------------------------
\bibliographystyle{plain}
\bibliography{normal_generated_py3}
% ----------------------------------------------------

% ----------------------------------------------------
\appendix
\section*{Appendix}
\label{section:Appendix}
% ----------------------------------------------------

% ----------------------------------------------------
\section{Performance Evaluation: Type I Attacks (GIN/GAT)}
\label{sec:type1_performance_appendix}
% ----------------------------------------------------

The performance of Type I attacks using GIN and GAT as surrogate models are summarized in \autoref{tab:acc_fidelity_type1_gin} and \autoref{tab:acc_fidelity_type1_gat}.

\begin{table*}[!t]
\centering
\caption{The accuracy and fidelity score of Type I attacks using different response information on all the 6 datasets. 
Both average values and standard deviations are reported. 
The accuracy differences (in parenthesis) of the surrogate models to the target models are also reported. 
We use GIN as the surrogate model. }
\scalebox{0.8}{
\begin{tabular}{lccccccc}
\toprule
\multirow{3}{*}{\textbf{Dataset}} &
  \multirow{3}{*}{\begin{tabular}[x]{@{}c@{}}$\mathcal{M}_S$\\(GIN)\end{tabular}} &
  \multicolumn{6}{c}{$\mathcal{M}_T$} \\
\cmidrule{3-8}
 &
   &
  \multicolumn{2}{c}{\textbf{GIN}} &
  \multicolumn{2}{c}{\textbf{GAT}} &
  \multicolumn{2}{c}{\textbf{SAGE}} \\
 \cmidrule(lr){3-4} \cmidrule(lr){5-6} \cmidrule(lr){7-8}
 &
   &
  \textbf{Accuracy} &
  \textbf{Fidelity} &
  \textbf{Accuracy} &
  \textbf{Fidelity} &
  \textbf{Accuracy} &
  \textbf{Fidelity} \\

\midrule
              &  Projection  &  0.710$\pm$0.028 (-0.162) & 0.737$\pm$0.031 & 0.677$\pm$0.005 (-0.161) & 0.696$\pm$0.008 & 0.719$\pm$0.033 (-0.139) & 0.761$\pm$0.037 \\
              &  Prediction  &  0.751$\pm$0.003 (-0.121) & 0.788$\pm$0.008 & 0.775$\pm$0.004 (-0.063) & 0.826$\pm$0.004 & 0.819$\pm$0.002 (-0.039) & 0.897$\pm$0.001 \\
\multirow{-3}{*}{\textbf{DBLP}}  &  Embedding  &  0.743$\pm$0.005 (-0.129) & 0.769$\pm$0.003 & 0.792$\pm$0.004 (-0.046) & 0.836$\pm$0.004 & 0.805$\pm$0.003 (-0.053) & 0.874$\pm$0.007 \\
\midrule
              &  Projection  &  0.838$\pm$0.027 (-0.086) & 0.861$\pm$0.030 & 0.753$\pm$0.015 (-0.152) & 0.767$\pm$0.015 & 0.776$\pm$0.058 (-0.133) & 0.813$\pm$0.066 \\
              &  Prediction  &  0.875$\pm$0.004 (-0.049) & 0.905$\pm$0.005 & 0.868$\pm$0.004 (-0.037) & 0.896$\pm$0.005 & 0.876$\pm$0.003 (-0.033) & 0.932$\pm$0.002 \\
\multirow{-3}{*}{\textbf{Pubmed}}  &  Embedding  &  0.883$\pm$0.004 (-0.041) & 0.912$\pm$0.003 & 0.877$\pm$0.004 (-0.028) & 0.903$\pm$0.003 & 0.884$\pm$0.001 (-0.025) & 0.937$\pm$0.004 \\
\midrule
              &  Projection  &  0.675$\pm$0.018 (-0.236) & 0.658$\pm$0.016 & 0.715$\pm$0.012 (-0.196) & 0.701$\pm$0.014 & 0.733$\pm$0.011 (-0.186) & 0.727$\pm$0.009 \\
              &  Prediction  &  0.808$\pm$0.005 (-0.103) & 0.816$\pm$0.006 & 0.878$\pm$0.004 (-0.033) & 0.909$\pm$0.002 & 0.895$\pm$0.001 (-0.024) & 0.942$\pm$0.002 \\
\multirow{-3}{*}{\textbf{Citeseer}}  &  embedding  &  0.791$\pm$0.005 (-0.120) & 0.806$\pm$0.007 & 0.889$\pm$0.005 (-0.022) & 0.913$\pm$0.005 & 0.881$\pm$0.006 (-0.038) & 0.917$\pm$0.006 \\
\midrule
              &  Projection  &  0.824$\pm$0.038 (-0.129) & 0.826$\pm$0.039 & 0.813$\pm$0.017 (-0.152) & 0.816$\pm$0.018 & 0.796$\pm$0.049 (-0.160) & 0.806$\pm$0.051 \\
              &  Prediction  &  0.920$\pm$0.002 (-0.033) & 0.925$\pm$0.003 & 0.945$\pm$0.001 (-0.020) & 0.948$\pm$0.002 & 0.956$\pm$0.002 (-0.000) & 0.976$\pm$0.001 \\
\multirow{-3}{*}{\textbf{Coauthor}}  &  Embedding  &  0.900$\pm$0.005 (-0.053) & 0.907$\pm$0.007 & 0.944$\pm$0.001 (-0.021) & 0.945$\pm$0.001 & 0.947$\pm$0.007 (-0.009) & 0.962$\pm$0.007 \\
\midrule
              &  Projection  &  0.824$\pm$0.020 (-0.105) & 0.819$\pm$0.023 & 0.835$\pm$0.044 (-0.100) & 0.849$\pm$0.048 & 0.892$\pm$0.018 (-0.045) & 0.913$\pm$0.022 \\
              &  Prediction  &  0.829$\pm$0.014 (-0.100) & 0.826$\pm$0.015 & 0.916$\pm$0.008 (-0.019) & 0.925$\pm$0.009 & 0.918$\pm$0.007 (-0.019) & 0.946$\pm$0.004 \\
\multirow{-3}{*}{\textbf{ACM}}  &  Embedding  &  0.860$\pm$0.010 (-0.069) & 0.859$\pm$0.019 & 0.920$\pm$0.006 (-0.015) & 0.930$\pm$0.004 & 0.917$\pm$0.014 (-0.020) & 0.944$\pm$0.014 \\
\midrule
              &  Projection  &  0.717$\pm$0.037 (-0.139) & 0.726$\pm$0.024 & 0.719$\pm$0.021 (-0.234) & 0.725$\pm$0.020 & 0.764$\pm$0.028 (-0.173) & 0.779$\pm$0.030 \\
              &  Prediction  &  0.893$\pm$0.006 (0.037) & 0.848$\pm$0.016 & 0.930$\pm$0.008 (-0.023) & 0.935$\pm$0.007 & 0.933$\pm$0.004 (-0.004) & 0.943$\pm$0.005 \\
\multirow{-3}{*}{\textbf{Amazon}}  &  Embedding  &  0.897$\pm$0.005 (0.041) & 0.830$\pm$0.021 & 0.929$\pm$0.006 (-0.024) & 0.935$\pm$0.006 & 0.931$\pm$0.005 (-0.006) & 0.934$\pm$0.006 \\
\bottomrule
\end{tabular}
}
\label{tab:acc_fidelity_type1_gin}
\end{table*}

\begin{table*}[!t]
\centering
\caption{The accuracy and fidelity score of Type I attacks using different response information on all the 6 datasets. 
Both average values and standard deviations are reported. 
The accuracy differences (in parenthesis) of the surrogate models to the target models are also reported. 
We use GAT as the surrogate model.}
\scalebox{0.8}{
\begin{tabular}{lccccccc}
\toprule
\multirow{3}{*}{\textbf{Dataset}} &
  \multirow{3}{*}{\begin{tabular}[x]{@{}c@{}}$\mathcal{M}_S$\\(GAT)\end{tabular}} &
  \multicolumn{6}{c}{$\mathcal{M}_T$} \\
\cmidrule{3-8}
 &
   &
  \multicolumn{2}{c}{\textbf{GIN}} &
  \multicolumn{2}{c}{\textbf{GAT}} &
  \multicolumn{2}{c}{\textbf{SAGE}} \\
 \cmidrule(lr){3-4} \cmidrule(lr){5-6} \cmidrule(lr){7-8}
 &
   &
  \textbf{Accuracy} &
  \textbf{Fidelity} &
  \textbf{Accuracy} &
  \textbf{Fidelity} &
  \textbf{Accuracy} &
  \textbf{Fidelity} \\
\midrule
              &  Projection  &  0.723$\pm$0.007 (-0.149) & 0.754$\pm$0.008 & 0.671$\pm$0.010 (-0.167) & 0.689$\pm$0.013 & 0.746$\pm$0.007 (-0.112) & 0.793$\pm$0.005 \\
              &  Prediction  &  0.745$\pm$0.006 (-0.127) & 0.784$\pm$0.008 & 0.782$\pm$0.004 (-0.056) & 0.834$\pm$0.005 & 0.809$\pm$0.005 (-0.049) & 0.880$\pm$0.005 \\
\multirow{-3}{*}{\textbf{DBLP}}  &  Embedding  &  0.696$\pm$0.010 (-0.176) & 0.722$\pm$0.007 & 0.796$\pm$0.006 (-0.042) & 0.844$\pm$0.006 & 0.762$\pm$0.039 (-0.096) & 0.819$\pm$0.048 \\
\midrule
              &  Projection  &  0.852$\pm$0.003 (-0.072) & 0.887$\pm$0.003 & 0.786$\pm$0.026 (-0.119) & 0.805$\pm$0.025 & 0.847$\pm$0.004 (-0.062) & 0.893$\pm$0.005 \\
              &  Prediction  &  0.862$\pm$0.000 (-0.062) & 0.899$\pm$0.002 & 0.862$\pm$0.002 (-0.043) & 0.899$\pm$0.003 & 0.863$\pm$0.004 (-0.046) & 0.921$\pm$0.004 \\
\multirow{-3}{*}{\textbf{Pubmed}}  &  Embedding  &  0.848$\pm$0.006 (-0.076) & 0.873$\pm$0.006 & 0.869$\pm$0.002 (-0.036) & 0.906$\pm$0.003 & 0.864$\pm$0.012 (-0.045) & 0.911$\pm$0.018 \\
\midrule
              &  Projection  &  0.743$\pm$0.008 (-0.168) & 0.746$\pm$0.007 & 0.755$\pm$0.003 (-0.156) & 0.763$\pm$0.009 & 0.786$\pm$0.009 (-0.133) & 0.798$\pm$0.011 \\
              &  Prediction  &  0.798$\pm$0.008 (-0.113) & 0.808$\pm$0.013 & 0.880$\pm$0.004 (-0.031) & 0.908$\pm$0.007 & 0.894$\pm$0.004 (-0.025) & 0.935$\pm$0.005 \\
\multirow{-3}{*}{\textbf{Citeseer}}  &  Embedding  &  0.776$\pm$0.013 (-0.135) & 0.783$\pm$0.012 & 0.895$\pm$0.002 (-0.016) & 0.929$\pm$0.005 & 0.885$\pm$0.004 (-0.034) & 0.912$\pm$0.004 \\
\midrule
              &  Projection  &  0.847$\pm$0.009 (-0.106) & 0.857$\pm$0.009 & 0.845$\pm$0.009 (-0.120) & 0.849$\pm$0.009 & 0.728$\pm$0.011 (-0.228) & 0.738$\pm$0.012 \\
              &  Prediction  &  0.879$\pm$0.006 (-0.074) & 0.889$\pm$0.006 & 0.932$\pm$0.004 (-0.033) & 0.939$\pm$0.004 & 0.889$\pm$0.009 (-0.067) & 0.903$\pm$0.009 \\
\multirow{-3}{*}{\textbf{Coauthor}}  &  Embedding  &  0.888$\pm$0.007 (-0.065) & 0.893$\pm$0.007 & 0.945$\pm$0.002 (-0.020) & 0.950$\pm$0.003 & 0.885$\pm$0.031 (-0.071) & 0.895$\pm$0.033 \\
\midrule
              &  Projection  &  0.848$\pm$0.022 (-0.081) & 0.848$\pm$0.021 & 0.871$\pm$0.014 (-0.064) & 0.895$\pm$0.011 & 0.893$\pm$0.019 (-0.044) & 0.922$\pm$0.019 \\
              &  Prediction  &  0.840$\pm$0.023 (-0.089) & 0.846$\pm$0.025 & 0.906$\pm$0.004 (-0.029) & 0.929$\pm$0.007 & 0.893$\pm$0.010 (-0.044) & 0.927$\pm$0.008 \\
\multirow{-3}{*}{\textbf{ACM}}  &  Embedding  &  0.869$\pm$0.012 (-0.060) & 0.878$\pm$0.009 & 0.902$\pm$0.011 (-0.033) & 0.922$\pm$0.009 & 0.870$\pm$0.024 (-0.067) & 0.896$\pm$0.024 \\
\midrule
              &  Projection  &  0.634$\pm$0.050 (-0.222) & 0.623$\pm$0.069 & 0.780$\pm$0.034 (-0.173) & 0.786$\pm$0.038 & 0.781$\pm$0.010 (-0.156) & 0.792$\pm$0.011 \\
              &  Prediction  &  0.804$\pm$0.030 (-0.052) & 0.780$\pm$0.051 & 0.921$\pm$0.006 (-0.032) & 0.937$\pm$0.006 & 0.893$\pm$0.013 (-0.044) & 0.907$\pm$0.017 \\
\multirow{-3}{*}{\textbf{Amazon}}  &  Embedding  &  0.857$\pm$0.012 (0.001) & 0.819$\pm$0.034 & 0.922$\pm$0.004 (-0.031) & 0.938$\pm$0.004 & 0.906$\pm$0.020 (-0.031) & 0.924$\pm$0.024 \\
\bottomrule
\end{tabular}
}
\label{tab:acc_fidelity_type1_gat}
\end{table*}

% ----------------------------------------------------
\section{Performance Evaluation: Type II Attacks (GIN/GAT)}
\label{sec:type2_performance_appendix}
% ----------------------------------------------------

The performance of Type II attacks using GIN and GAT as surrogate models are summarized in \autoref{tab:acc_fidelity_type2_gin} and \autoref{tab:acc_fidelity_type2_gat}.

\begin{table*}[!t]
\centering
\caption{The accuracy and fidelity score of Type II attacks using different response information on all the 6 datasets. 
Both average values and standard deviations are reported. 
The accuracy differences (in parenthesis) of the surrogate models to the target models are also reported. 
We use GIN as the surrogate model.}
\scalebox{0.8}{
\begin{tabular}{lccccccc}
\toprule
\multirow{3}{*}{\textbf{Dataset}} &
  \multirow{3}{*}{\begin{tabular}[x]{@{}c@{}}$\mathcal{M}_S$\\(GIN)\end{tabular}} &
  \multicolumn{6}{c}{$\mathcal{M}_T$} \\
\cmidrule{3-8}
 &
  &
  \multicolumn{2}{c}{\textbf{GIN}} &
  \multicolumn{2}{c}{\textbf{GAT}} &
  \multicolumn{2}{c}{\textbf{SAGE}} \\
 \cmidrule(lr){3-4} \cmidrule(lr){5-6} \cmidrule(lr){7-8}
 &
  &
  \textbf{Accuracy} &
  \textbf{Fidelity} &
  \textbf{Accuracy} &
  \textbf{Fidelity} &
  \textbf{Accuracy} &
  \textbf{Fidelity} \\

\midrule
              &  Projection  &  0.715$\pm$0.028 (-0.157) & 0.751$\pm$0.035 & 0.680$\pm$0.005 (-0.158) & 0.701$\pm$0.006 & 0.724$\pm$0.025 (-0.134) & 0.775$\pm$0.029 \\
              &  Prediction  &  0.788$\pm$0.002 (-0.084) & 0.839$\pm$0.005 & 0.775$\pm$0.003 (-0.063) & 0.835$\pm$0.003 & 0.818$\pm$0.004 (-0.040) & 0.899$\pm$0.006 \\
\multirow{-3}{*}{\textbf{DBLP}}  &  Embedding  &  0.785$\pm$0.012 (-0.087) & 0.826$\pm$0.015 & 0.794$\pm$0.004 (-0.044) & 0.846$\pm$0.015 & 0.803$\pm$0.004 (-0.055) & 0.876$\pm$0.004 \\
\midrule
              &  Projection  &  0.826$\pm$0.029 (-0.098) & 0.851$\pm$0.034 & 0.736$\pm$0.022 (-0.169) & 0.750$\pm$0.020 & 0.818$\pm$0.064 (-0.091) & 0.855$\pm$0.076 \\
              &  Prediction  &  0.873$\pm$0.004 (-0.051) & 0.904$\pm$0.003 & 0.870$\pm$0.003 (-0.035) & 0.898$\pm$0.002 & 0.876$\pm$0.002 (-0.033) & 0.933$\pm$0.005 \\
\multirow{-3}{*}{\textbf{Pubmed}}  &  Embedding  &  0.885$\pm$0.004 (-0.039) & 0.913$\pm$0.004 & 0.882$\pm$0.001 (-0.023) & 0.910$\pm$0.003 & 0.882$\pm$0.004 (-0.027) & 0.933$\pm$0.007 \\
\midrule
              &  Projection  &  0.665$\pm$0.026 (-0.246) & 0.673$\pm$0.023 & 0.673$\pm$0.021 (-0.238) & 0.679$\pm$0.014 & 0.726$\pm$0.030 (-0.193) & 0.740$\pm$0.026 \\
              &  Prediction  &  0.841$\pm$0.014 (-0.070) & 0.877$\pm$0.010 & 0.880$\pm$0.008 (-0.031) & 0.922$\pm$0.003 & 0.896$\pm$0.005 (-0.023) & 0.952$\pm$0.006 \\
\multirow{-3}{*}{\textbf{Citeseer}}  &  Embedding  &  0.814$\pm$0.014 (-0.097) & 0.836$\pm$0.012 & 0.889$\pm$0.005 (-0.022) & 0.924$\pm$0.004 & 0.884$\pm$0.004 (-0.035) & 0.930$\pm$0.003 \\
\midrule
              &  Projection  &  0.859$\pm$0.040 (-0.094) & 0.849$\pm$0.040 & 0.839$\pm$0.053 (-0.126) & 0.844$\pm$0.054 & 0.744$\pm$0.048 (-0.212) & 0.751$\pm$0.048 \\
              &  Prediction  &  0.951$\pm$0.001 (-0.002) & 0.955$\pm$0.001 & 0.955$\pm$0.003 (-0.010) & 0.960$\pm$0.002 & 0.956$\pm$0.001 (-0.000) & 0.979$\pm$0.001 \\
\multirow{-3}{*}{\textbf{Coauthor}}  &  Embedding  &  0.947$\pm$0.004 (-0.006) & 0.953$\pm$0.003 & 0.951$\pm$0.001 (-0.014) & 0.954$\pm$0.002 & 0.949$\pm$0.004 (-0.007) & 0.963$\pm$0.006 \\
\midrule
              &  Projection  &  0.856$\pm$0.036 (-0.073) & 0.850$\pm$0.047 & 0.876$\pm$0.055 (-0.059) & 0.886$\pm$0.056 & 0.893$\pm$0.030 (-0.044) & 0.921$\pm$0.030 \\
              &  Prediction  &  0.902$\pm$0.008 (-0.027) & 0.899$\pm$0.012 & 0.928$\pm$0.005 (-0.007) & 0.941$\pm$0.006 & 0.934$\pm$0.006 (-0.003) & 0.962$\pm$0.004 \\
\multirow{-3}{*}{\textbf{ACM}}  &  Embedding  &  0.901$\pm$0.011 (-0.028) & 0.891$\pm$0.017 & 0.925$\pm$0.005 (-0.010) & 0.937$\pm$0.004 & 0.928$\pm$0.007 (-0.009) & 0.947$\pm$0.008 \\
\midrule
              &  Projection  &  0.653$\pm$0.052 (-0.203) & 0.668$\pm$0.058 & 0.657$\pm$0.039 (-0.296) & 0.656$\pm$0.041 & 0.764$\pm$0.029 (-0.173) & 0.774$\pm$0.028 \\
              &  Prediction  &  0.892$\pm$0.018 (0.036) & 0.854$\pm$0.016 & 0.907$\pm$0.010 (-0.046) & 0.914$\pm$0.005 & 0.943$\pm$0.005 (0.006) & 0.953$\pm$0.006 \\
\multirow{-3}{*}{\textbf{Amazon}}  &  Embedding  &  0.912$\pm$0.010 (0.056) & 0.860$\pm$0.006 & 0.933$\pm$0.005 (-0.020) & 0.941$\pm$0.010 & 0.939$\pm$0.002 (0.002) & 0.956$\pm$0.004 \\
\bottomrule
\end{tabular}
}
\label{tab:acc_fidelity_type2_gin}
\end{table*}

\begin{table*}[!t]
\centering
\caption{The accuracy and fidelity score of Type II attacks using different response information on all the 6 datasets. 
Both average values and standard deviations are reported. 
The accuracy differences (in parenthesis) of the surrogate models to the target models are also reported. 
We use GAT as the surrogate model.}
\scalebox{0.8}{
\begin{tabular}{lccccccc}
\toprule
\multirow{3}{*}{\textbf{Dataset}} &
  \multirow{3}{*}{\begin{tabular}[x]{@{}c@{}}$\mathcal{M}_S$\\(GAT)\end{tabular}} &
  \multicolumn{6}{c}{$\mathcal{M}_T$} \\
\cmidrule{3-8}
 &
  &
  \multicolumn{2}{c}{\textbf{GIN}} &
  \multicolumn{2}{c}{\textbf{GAT}} &
  \multicolumn{2}{c}{\textbf{SAGE}} \\
 \cmidrule(lr){3-4} \cmidrule(lr){5-6} \cmidrule(lr){7-8}
 &
  &
  \textbf{Accuracy} &
  \textbf{Fidelity} &
  \textbf{Accuracy} &
  \textbf{Fidelity} &
  \textbf{Accuracy} &
  \textbf{Fidelity} \\

\midrule

              &  Projection  &  0.702$\pm$0.023 (-0.170) & 0.727$\pm$0.027 & 0.668$\pm$0.007 (-0.170) & 0.693$\pm$0.011 & 0.746$\pm$0.023 (-0.112) & 0.793$\pm$0.032 \\
              &  Prediction  &  0.739$\pm$0.010 (-0.133) & 0.784$\pm$0.012 & 0.779$\pm$0.005 (-0.059) & 0.834$\pm$0.014 & 0.806$\pm$0.004 (-0.052) & 0.881$\pm$0.005 \\
\multirow{-3}{*}{\textbf{DBLP}}  &  Embedding  &  0.741$\pm$0.006 (-0.131) & 0.774$\pm$0.009 & 0.800$\pm$0.006 (-0.038) & 0.853$\pm$0.008 & 0.743$\pm$0.032 (-0.115) & 0.779$\pm$0.042 \\
\midrule
              &  Projection  &  0.822$\pm$0.023 (-0.102) & 0.850$\pm$0.024 & 0.746$\pm$0.054 (-0.159) & 0.765$\pm$0.053 & 0.844$\pm$0.024 (-0.065) & 0.889$\pm$0.030 \\
              &  Prediction  &  0.837$\pm$0.008 (-0.087) & 0.869$\pm$0.010 & 0.861$\pm$0.003 (-0.044) & 0.896$\pm$0.004 & 0.861$\pm$0.002 (-0.048) & 0.917$\pm$0.004 \\
\multirow{-3}{*}{\textbf{Pubmed}}  &  Embedding  &  0.854$\pm$0.011 (-0.070) & 0.880$\pm$0.012 & 0.873$\pm$0.004 (-0.032) & 0.911$\pm$0.003 & 0.863$\pm$0.006 (-0.046) & 0.909$\pm$0.010 \\
\midrule
              &  Projection  &  0.704$\pm$0.017 (-0.207) & 0.719$\pm$0.017 & 0.750$\pm$0.022 (-0.161) & 0.762$\pm$0.025 & 0.768$\pm$0.019 (-0.151) & 0.796$\pm$0.022 \\
              &  Prediction  &  0.800$\pm$0.010 (-0.111) & 0.838$\pm$0.014 & 0.883$\pm$0.005 (-0.028) & 0.916$\pm$0.006 & 0.890$\pm$0.002 (-0.029) & 0.941$\pm$0.003 \\
\multirow{-3}{*}{\textbf{Citeseer}}  &  Embedding  &  0.807$\pm$0.023 (-0.104) & 0.819$\pm$0.026 & 0.893$\pm$0.011 (-0.018) & 0.932$\pm$0.007 & 0.876$\pm$0.005 (-0.043) & 0.914$\pm$0.007 \\
\midrule
              &  Projection  &  0.643$\pm$0.100 (-0.310) & 0.649$\pm$0.101 & 0.762$\pm$0.077 (-0.203) & 0.764$\pm$0.080 & 0.724$\pm$0.031 (-0.232) & 0.733$\pm$0.031 \\
              &  Prediction  &  0.823$\pm$0.052 (-0.130) & 0.831$\pm$0.054 & 0.871$\pm$0.033 (-0.094) & 0.877$\pm$0.034 & 0.795$\pm$0.029 (-0.161) & 0.806$\pm$0.032 \\
\multirow{-3}{*}{\textbf{Coauthor}}  &  Embedding  &  0.864$\pm$0.020 (-0.089) & 0.871$\pm$0.021 & 0.882$\pm$0.036 (-0.083) & 0.887$\pm$0.036 & 0.741$\pm$0.085 (-0.215) & 0.750$\pm$0.087 \\
\midrule
              &  Projection  &  0.834$\pm$0.011 (-0.095) & 0.838$\pm$0.009 & 0.896$\pm$0.012 (-0.039) & 0.921$\pm$0.011 & 0.910$\pm$0.008 (-0.027) & 0.940$\pm$0.011 \\
              &  Prediction  &  0.880$\pm$0.004 (-0.049) & 0.889$\pm$0.021 & 0.915$\pm$0.007 (-0.020) & 0.946$\pm$0.007 & 0.917$\pm$0.004 (-0.020) & 0.956$\pm$0.004 \\
\multirow{-3}{*}{\textbf{ACM}}  &  Embedding  &  0.909$\pm$0.006 (-0.020) & 0.913$\pm$0.012 & 0.907$\pm$0.007 (-0.028) & 0.928$\pm$0.008 & 0.908$\pm$0.015 (-0.029) & 0.941$\pm$0.018 \\
\midrule
              &  Projection  &  0.674$\pm$0.040 (-0.182) & 0.590$\pm$0.058 & 0.661$\pm$0.055 (-0.292) & 0.557$\pm$0.060 & 0.739$\pm$0.015 (-0.198) & 0.694$\pm$0.019 \\
              &  Prediction  &  0.765$\pm$0.076 (-0.091) & 0.653$\pm$0.074 & 0.881$\pm$0.024 (-0.072) & 0.753$\pm$0.022 & 0.849$\pm$0.024 (-0.088) & 0.752$\pm$0.029 \\
\multirow{-3}{*}{\textbf{Amazon}}  &  Embedding  &  0.853$\pm$0.035 (-0.003) & 0.767$\pm$0.023 & 0.897$\pm$0.010 (-0.056) & 0.869$\pm$0.014 & 0.832$\pm$0.043 (-0.105) & 0.756$\pm$0.051 \\
\bottomrule
\end{tabular}
}
\label{tab:acc_fidelity_type2_gat}
\end{table*}

% ----------------------------------------------------
\section{Fidelity}
\label{sec:random_knn_idgl_comparison_fidelity}
% ----------------------------------------------------

\begin{figure}[t]
\centering
\includegraphics[width=0.8\linewidth]{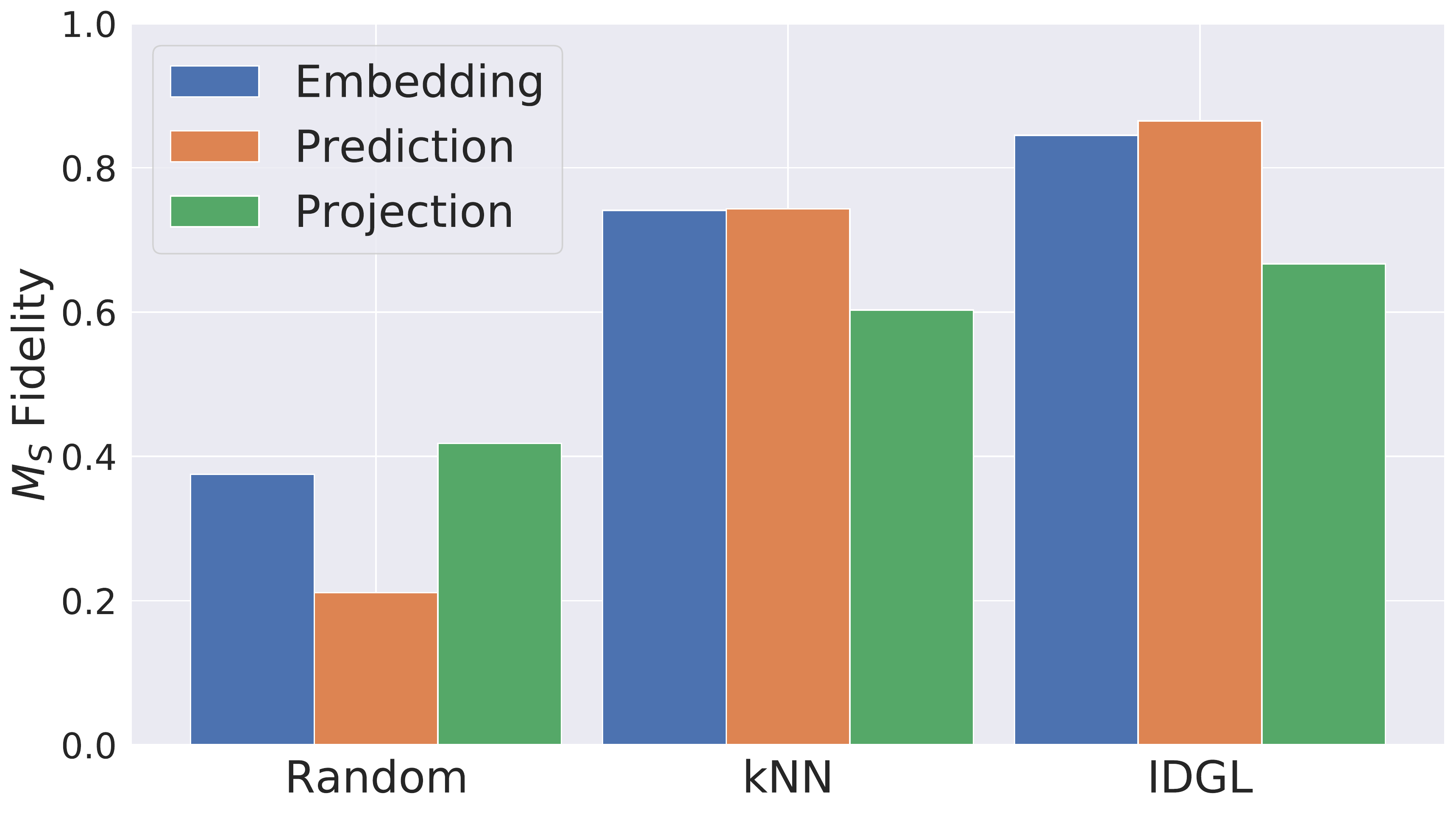}
\caption{The fidelity score of Type II attacks on the Citeseer dataset using different graph reconstruction methods. 
We use GAT as the target model and GraphSAGE as the surrogate model.}
\label{fig:graph_reconstruction_comparison_fidelity}
\end{figure}

The fidelity score of Type II attacks on the Citeseer dataset using different graph reconstruction methods is shown in \autoref{fig:graph_reconstruction_comparison_fidelity}.

% ----------------------------------------------------
\section{Hyperparameter Study}
\label{sec:hyperparams}
% ----------------------------------------------------

\begin{figure}[!t]
\centering
\begin{subfigure}[t]{0.49\linewidth}
\includegraphics[width=\textwidth]{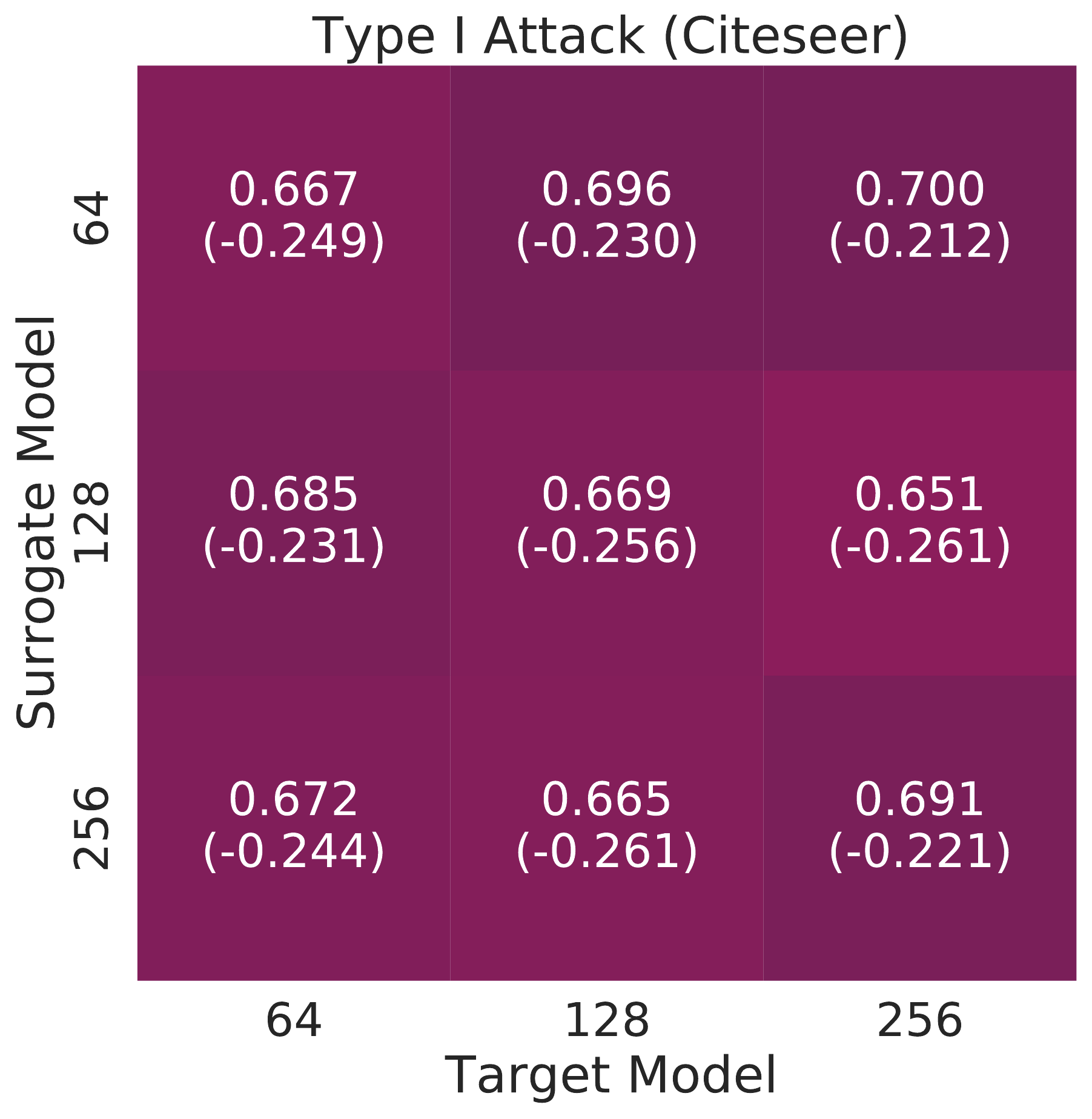}
\caption{t-SNE Projection} 
\label{fig:citeseer_hiddenunit_projection_sage}
\end{subfigure}
\begin{subfigure}[t]{0.49\linewidth}
\includegraphics[width=\textwidth]{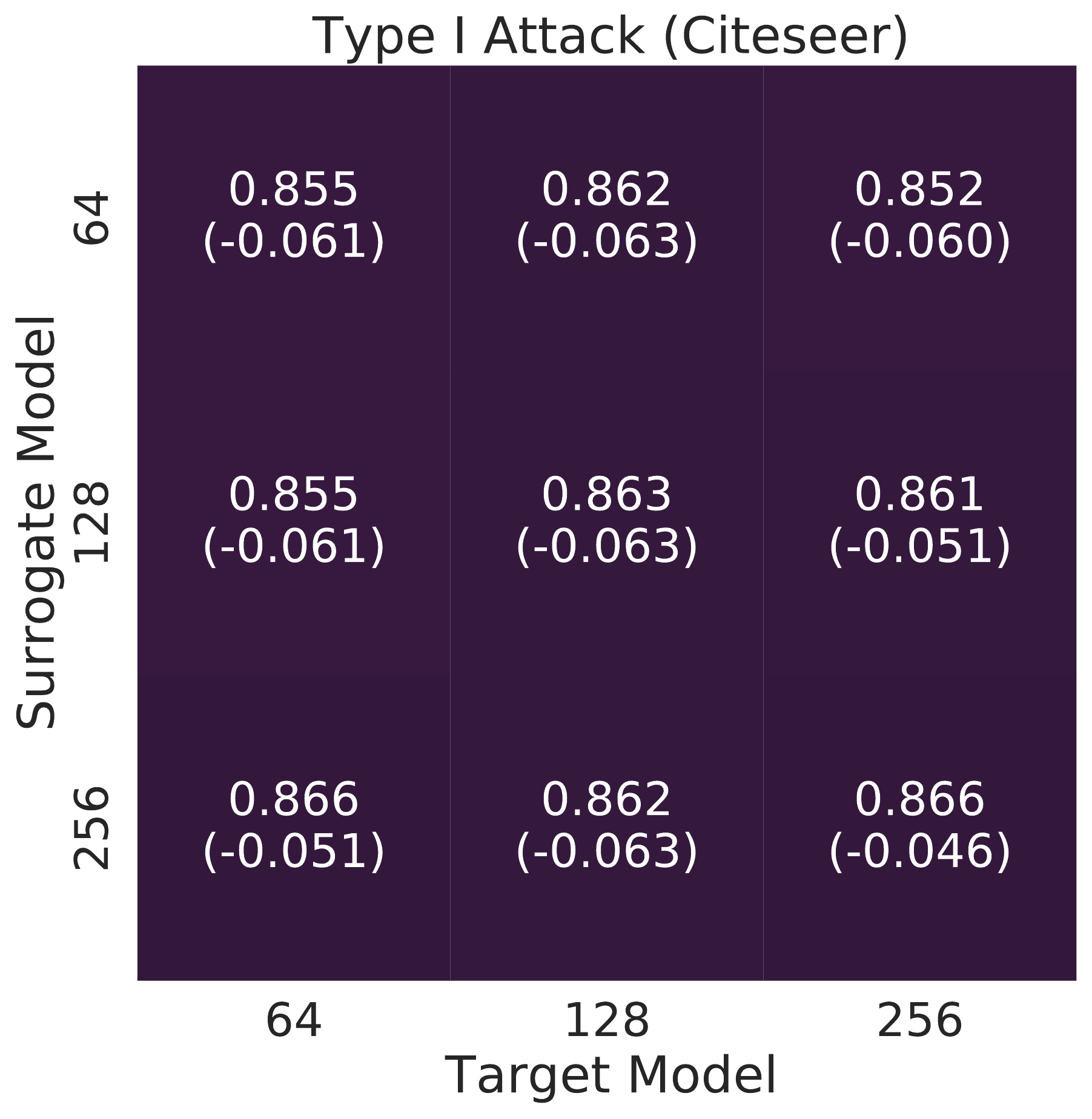}
\caption{Prediction} 
\label{fig:citeseer_hiddenunit_prediction_sage}
\end{subfigure}
\caption{Hyperparameter study: hidden unit size.
We show the influence of different combinations of the hidden unit size on the surrogate model's performance.}
\end{figure}

\begin{figure}[!t]
\centering
\begin{subfigure}[t]{0.49\linewidth}
\includegraphics[width=\textwidth]{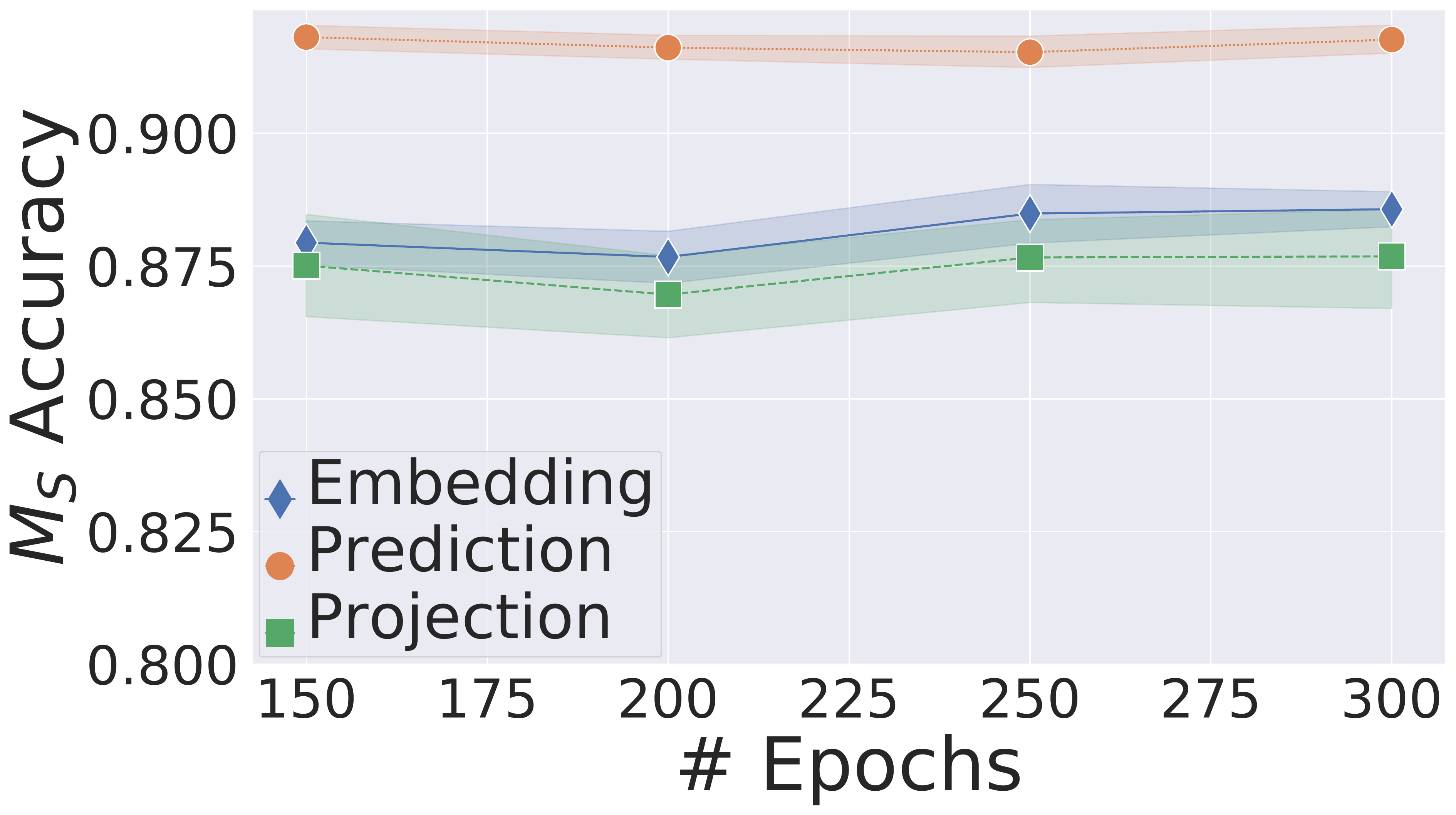}
\caption{Target model (GAT)} 
\label{fig:acm_GAT_epoch_hyper}
\end{subfigure}
\centering
\begin{subfigure}[t]{0.49\linewidth}
\includegraphics[width=\textwidth]{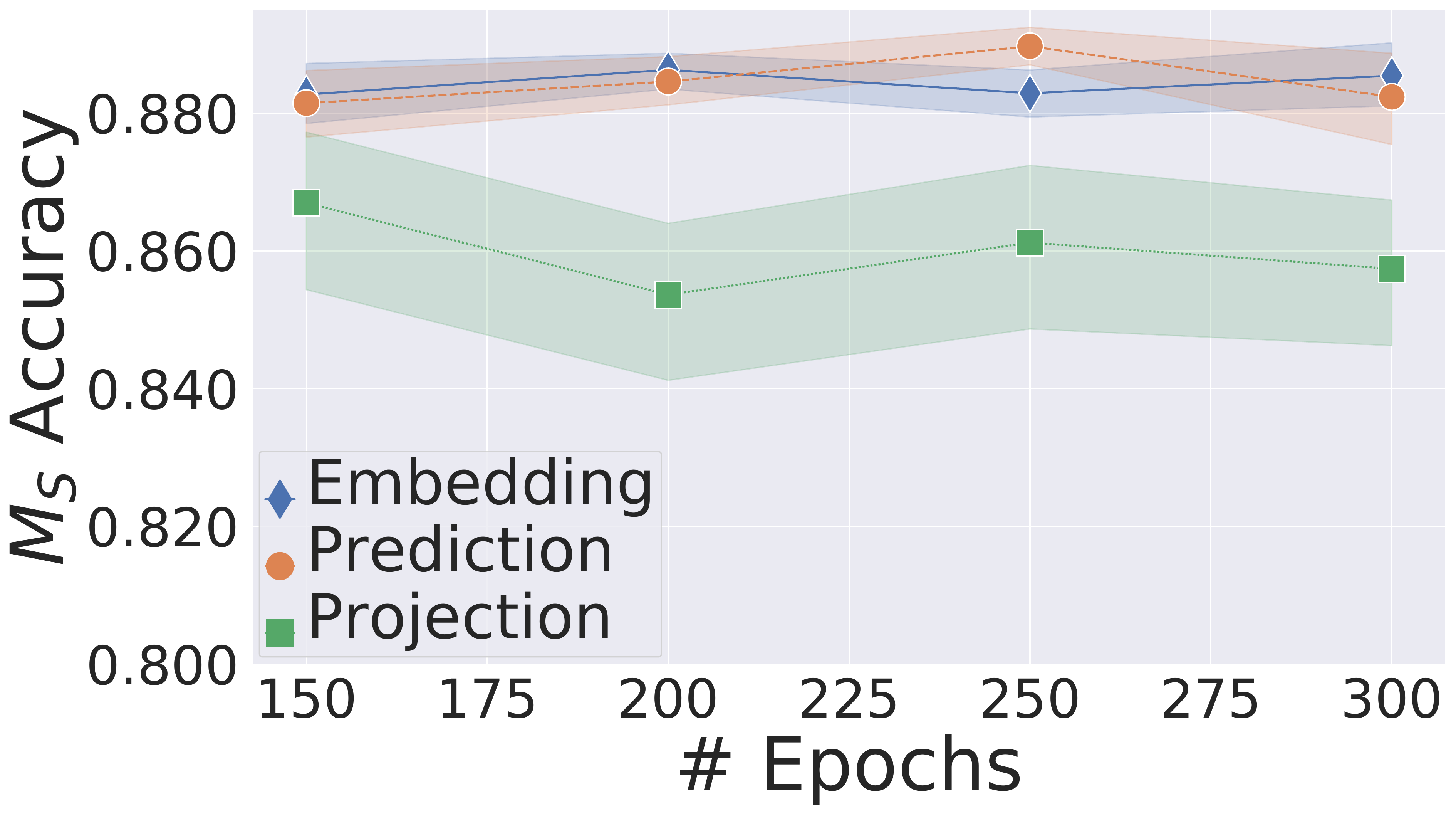}
\caption{Target model (GIN)} 
\label{fig:acm_GIN_epoch_hyper}
\end{subfigure}
\caption{Hyperparameter study: number of epochs. 
We report the accuracy and standard deviation of the surrogate model (GraphSAGE) under Type I attacks given different number of epochs. 
We fix the dataset to ACM.}
\label{fig:hyperparam_num_epochs}
\end{figure}

\begin{figure}[!t]
\centering
\begin{subfigure}[t]{0.49\linewidth}
\includegraphics[width=\textwidth]{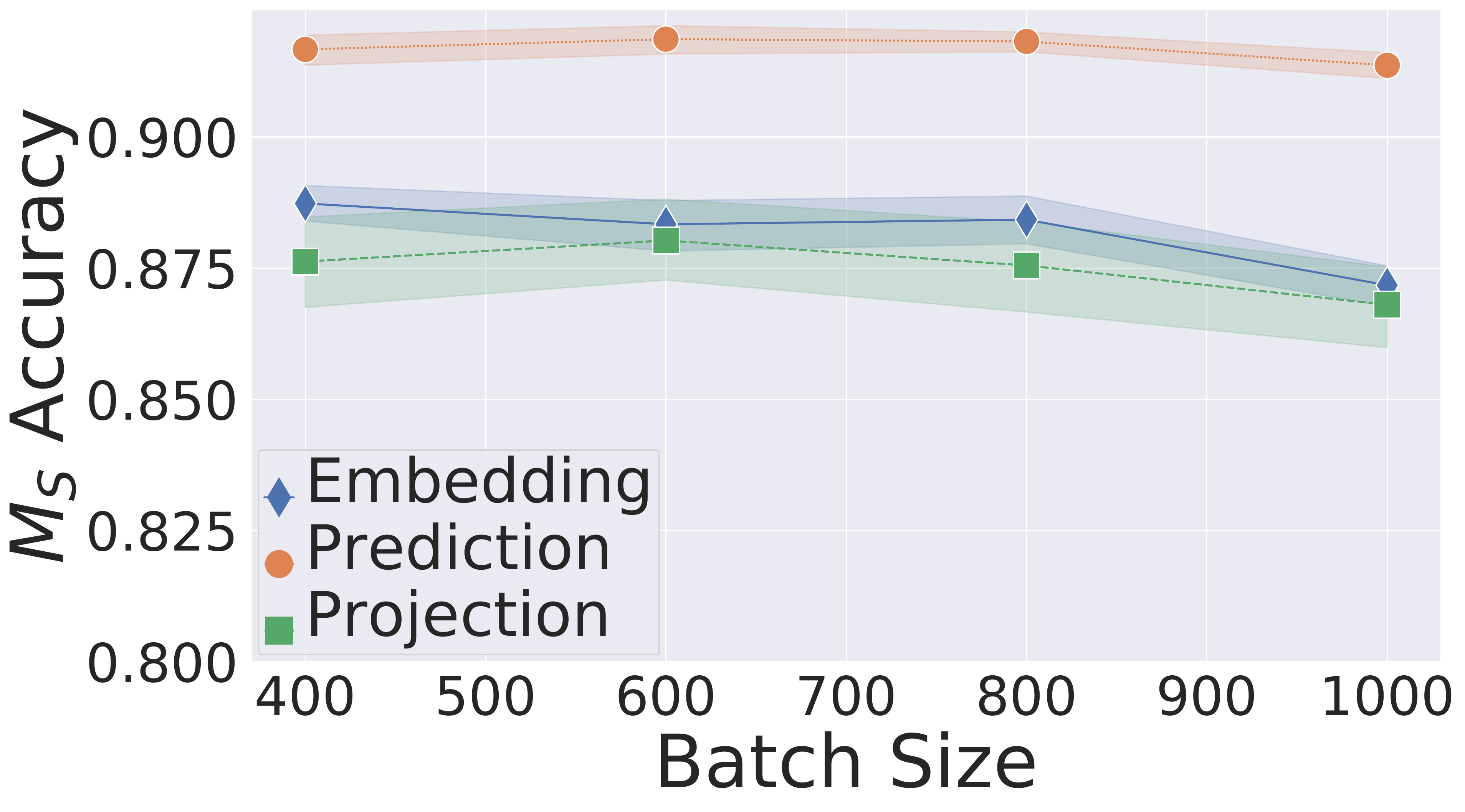}
\caption{Target model (GAT)} 
\label{fig:acm_GAT_batch_hyper}
\end{subfigure}
\centering
\begin{subfigure}[t]{0.49\linewidth}
\includegraphics[width=\textwidth]{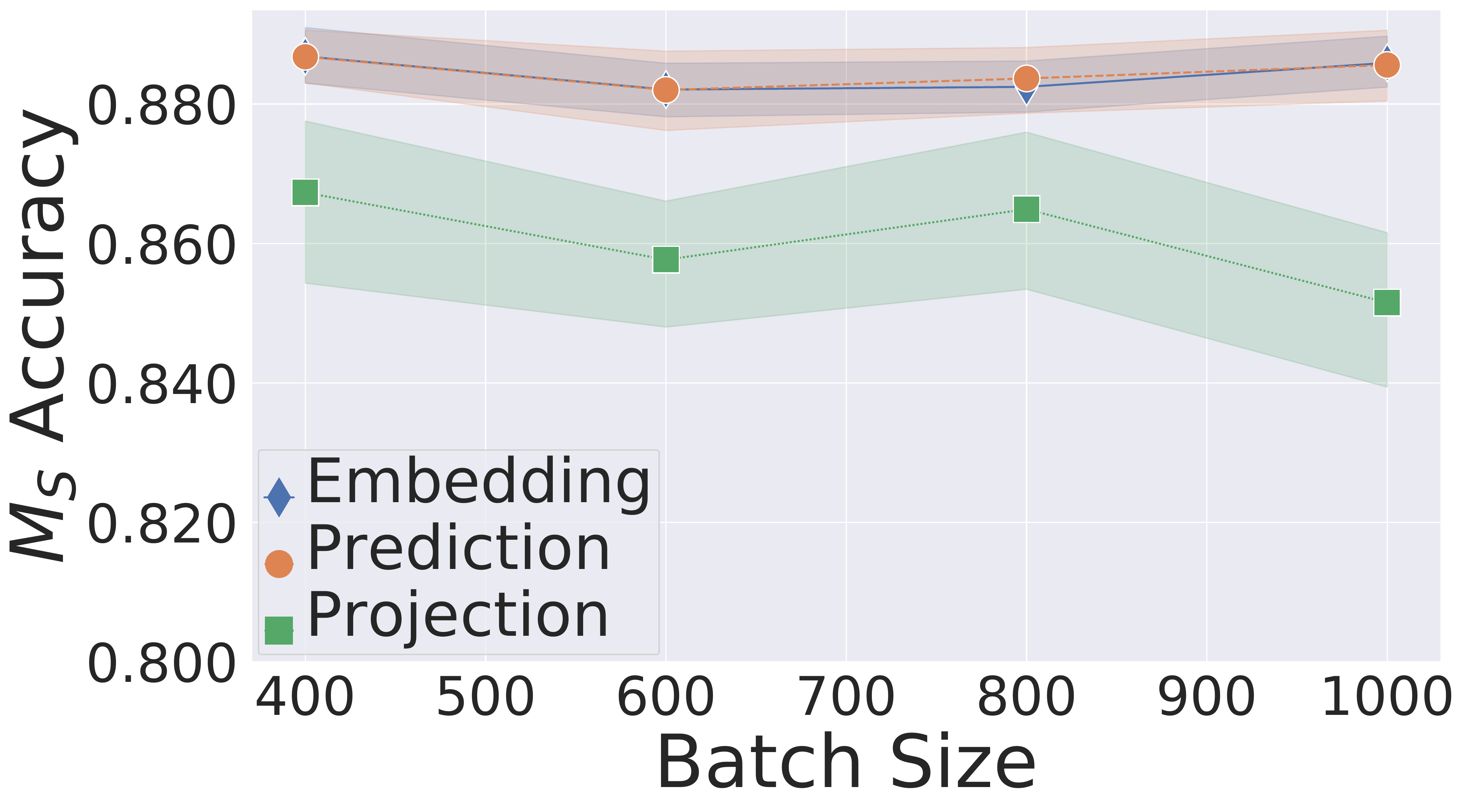}
\caption{Target model (GIN)} 
\label{fig:acm_GIN_batch_hyper}
\end{subfigure}
\caption{Hyperparameter study: batch size. 
We report the accuracy and standard deviation of the surrogate model (GraphSAGE) under Type I attacks given different batch sizes. 
We fix the dataset to ACM.}
\label{fig:hyperparam_batch_size}
\end{figure}

GNNs are complex and their performance may be affected by the hyperparameter settings.
This is especially important for our model stealing attack since we use inductive GNNs as the part of our surrogate models.
At the same time, our attack is in a full black-box setting.
As such, we evaluate the impact of three hyperparameters on the performance of surrogate models, i.e., hidden unit size, number of epochs, and batch size.
For each hyperparameter, we only show the results on one dataset under Type I attacks given the space limitation.
Other datasets follow a similar trend.

\mypara{Hidden Unit Size}
In general, the larger the hidden units, the greater space of representation functions a graph convolutional layer can offer.
As such, we investigate how the hidden unit size may affect the attack performance.
Recall that the adversary does not know the target model's hidden unit size. 
They can only blindly guess the hidden unit size used by the target model. 
To this end, we first build a target model (i.e., GAT) with 64, 128, and 256 hidden units using Citeseer dataset.
We then launch Type I attacks using a fixed surrogate model (i.e., GraphSAGE) with 64, 128, and 256 hidden units.
In this way, we can observe the potential performance changes in different circumstances. 
The results are shown in \autoref{fig:citeseer_hiddenunit_projection_sage} and \autoref{fig:citeseer_hiddenunit_prediction_sage}, respectively.
We can see that different hidden unit sizes adopted by the surrogate models have a limited impact on the accuracy performance. 
Our hypothesis is that the inductive GNN models employed as surrogate models are powerful enough to extract information from the responses.

\mypara{Number of Epochs and Batch Size}
The number of epochs and batch size are the other two hyperparameters that may affect attack performance but can be controlled by the adversary.
Respectively they control the number of complete passes through the training dataset and the number of samples processed before the GNN model is updated.
Batch size may affect the speed and stability of the learning process, while the number of epochs may lead to overfitting.
To this end, we fix the surrogate model to GraphSAGE and the target models to GIN and GAT to understand the impact of both hyperparameters.
We use 150, 200, 250, 300 for the number of epochs, and 400, 600, 800, 1000 for bath sizes.
The results are summarized in \autoref{fig:hyperparam_num_epochs} and \autoref{fig:hyperparam_batch_size}, respectively.
Regarding the number of epochs, the accuracy is relatively stable with respect to different numbers of epochs (see \autoref{fig:hyperparam_num_epochs}).
Meanwhile, as we can observe in \autoref{fig:hyperparam_batch_size}, a larger batch size may have some negative impact on the attack accuracy.

\mypara{Takeaways}
Inductive GNN models employed as surrogate models are powerful enough to extract information from the responses.
We observe that a large batch size may have a negative impact on the attack accuracy.
Other hyperparameters such as hidden unit size and the number of epochs have a limited impact on the accuracy of the surrogate models.

% ----------------------------------------------------
\end{document}

%% file: latex_config.tex
\usepackage{zhanggroup}
\usepackage{xspace}
\usepackage{cite}
\usepackage{amsmath,amssymb,amsfonts}
\usepackage{bm}
\usepackage{algorithmic}
\usepackage{graphicx}
\usepackage{textcomp}
\usepackage{xcolor}
\usepackage{tikz}
\usepackage{array}
\usepackage{pifont}
\usepackage[absolute]{textpos}
\usepackage{multirow}
\usepackage{subcaption}
\usepackage{caption}
\usepackage{booktabs}
\newcommand{\mypara}[1]{\smallskip\noindent\textbf{#1.}\xspace}